\documentclass[runningheads]{llncs}

\usepackage{eccv}

\usepackage{multirow,array}
\usepackage{color, colortbl}
\usepackage[percent]{overpic}
\usepackage{eccvabbrv}

\usepackage{graphicx}
\usepackage{booktabs}

\usepackage[accsupp]{axessibility}  
\usepackage{afterpage}

\usepackage[pagebackref,breaklinks,colorlinks,citecolor=eccvblue]{hyperref}

\usepackage{orcidlink}
\DeclareMathOperator*{\argmax}{arg\,max}

\newcommand{\decparams}{\boldsymbol\theta}
\newcommand{\encparams}{\boldsymbol\phi}
\newcommand{\supsection}[1]{Supp.\ \textit{Sec.\ #1}}

\definecolor{Gray}{gray}{0.9}

\def\eg{\emph{e.g}\onedot} 
\def\ie{\emph{i.e}\onedot} 
 
\def\etc{\emph{etc}\onedot} \def\vs{\emph{vs}\onedot}
\def\wrt{w.r.t\onedot}

\newcommand{\VAEs}{\mbox{\textsc{VAE}}s\xspace}

\newcommand{\HVAE}{\mbox{\textsc{HVAE}}\xspace}
\newcommand{\HVAEs}{\mbox{\textsc{HVAEs}}\xspace}

\newcommand{\HDN}{\mbox{\textsc{HDN}}\xspace}

\newcommand{\muSplit}{\mbox{$\mu\text{Split}$}\xspace}

\newcommand{\regularLC}{\mbox{\textit{Regular-LC}}\xspace}

\newcommand{\hdnusplit}{\mbox{\textit{HDN$\oplus$\muSplit}}\xspace}
\newcommand{\usplitKL}{\mbox{\textit{Altered \muSplit}}\xspace}
\newcommand{\denoiSplit}{\mbox{$\text{denoi}\mathbb{S}\text{plit}$}\xspace}

\newcommand\tabOverallPerf{
\begin{table}[]
    \centering
    \begin{tabular}{|c|l|c|c|c|c|c|c|c|c|c|}
    \hline
    \multirow{3}{*}{Task}&\multirow{3}{*}{Model} & \multirow{3}{*}{training} & \multicolumn{8}{c|}{Noise level parameters}\\\cline{4-11}   
    & & & \multicolumn{4}{c|}{$\lambda=0$} & \multicolumn{4}{c|}{$\lambda=1000$}\\\cline{4-11}      
    & & $[h]$ &$\sigma=1$ & 1.5 & 2 & 4 & $\sigma=1$ & 1.5 & 2 & 4\\
    \hline
    \multicolumn{6}{c}{} \\[-10pt]
    \hline
    \multirow{8}{*}{T1} &\multirow{2}{*}{\muSplit} &\multirow{2}{*}{7}&  30.3 & 28.4 &27.4 & 25.9& 29.4 & 28.1 &27.3 & 25.9\\
    &  & & \scriptsize{0.853} & \scriptsize{0.748} & \scriptsize{0.66} & \scriptsize{0.42} & \scriptsize{0.844} & \scriptsize{0.750} & \scriptsize{0.667} & \scriptsize{0.437} \\ \cline{2-11}
    & \multirow{2}{*}{\hdnusplit} &\multirow{2}{*}{11}& 37.3 & 34.9 & 33.8 & 29.4 & 36.3 & 34.3 & 33.3 & 29.4 \\
    &  & & \scriptsize{0.982} & \scriptsize{0.969} & \scriptsize{0.959}& \scriptsize{0.872}& \scriptsize{0.978} & \scriptsize{0.965} & \scriptsize{0.954} & \scriptsize{0.874}\\\cline{2-11}
    & \multirow{2}{*}{\usplitKL \textit{\textbf{(ours)}}} &\multirow{2}{*}{1.3}&  38.9 & 36.7 & 34.4 & 30.9 & 36.9 & 35.5 & 34.8 & 31.1\\
    &   && \scriptsize{0.988} & \textbf{\scriptsize{0.980}} & \scriptsize{0.965} & \scriptsize{0.909} & \scriptsize{0.982} &0.974 & \textbf{\scriptsize{0.968}} & \textbf{\scriptsize{0.912}}\\\cline{2-11}
    &\multirow{2}{*}{\denoiSplit \textit{\textbf{(ours)}}} &\multirow{2}{*}{1.5}& \textbf{39.7} & \textbf{36.8} & \textbf{35.4} & \textbf{31.1}& \textbf{37.9} & \textbf{36.3} & \textbf{35.0} & \textbf{31.2}\\
    &  & & \textbf{\scriptsize{0.989}} & \scriptsize{0.978} & \textbf{\scriptsize{0.969}} & \textbf{\scriptsize{0.912}} & \textbf{\scriptsize{0.984}} & \textbf{\scriptsize{0.977}} & \scriptsize{0.967} & \textbf{\scriptsize{0.912}}\\\hline
    \multicolumn{6}{c}{} \\[-10pt]
    \hline
    \multirow{8}{*}{T2} &\multirow{2}{*}{\muSplit} &\multirow{2}{*}{6.3}&  26.0 & 23.9 &22.7 &21.0& 25.2 & 23.7 &22.7 &21.1\\
    & & & \scriptsize{0.800} & \scriptsize{0.699} & \scriptsize{0.593} & \scriptsize{0.356} & \scriptsize{0.780} & \scriptsize{0.691} & \scriptsize{0.613} & \scriptsize{0.386} \\ \cline{2-11}
    &\multirow{2}{*}{\hdnusplit} &\multirow{2}{*}{10}& 30.1 & 28.4& 27.6& \textbf{25.3} & 29.6 & 28.4 & 27.4 & \textbf{25.2}\\
    & & & \scriptsize{0.909} & \scriptsize{0.873}& \scriptsize{0.845}& \textbf{\scriptsize{0.731}}& \textbf{\scriptsize{0.904}} & \scriptsize{0.874} & \scriptsize{0.835} & \textbf{\scriptsize{0.738}}\\ \cline{2-11}
    &\multirow{2}{*}{\usplitKL \textit{\textbf{(ours)}}} &\multirow{2}{*}{1.5}& 30.4 & 28.8 & 26.9 & 23.4 & \textbf{29.9} & 27.4 & \textbf{27.9} & 24.4\\
    & & & \scriptsize{0.915} & \scriptsize{0.879} & \scriptsize{0.809} & \scriptsize{0.620} & \scriptsize{0.903} & \scriptsize{0.833} & \textbf{\scriptsize{0.845}} & \scriptsize{0.677}\\ \cline{2-11}
    &\multirow{2}{*}{\denoiSplit \textit{\textbf{(ours)}}} &\multirow{2}{*}{1.6}& \textbf{30.5} & \textbf{29.2} & \textbf{28.2} & 25.1 & \textbf{29.9} & \textbf{29.0} & 27.0 & 24.8\\
    & & & \textbf{\scriptsize{0.916}} & \textbf{\scriptsize{0.886}} & \textbf{\scriptsize{0.860}} & \scriptsize{0.714} & \scriptsize{0.901} & \textbf{\scriptsize{0.885}} & \scriptsize{0.815} & \scriptsize{0.702}\\\hline
    \multicolumn{6}{c}{} \\[-10pt]
    \hline
    \multirow{8}{*}{T3} &\multirow{2}{*}{\muSplit} &\multirow{2}{*}{7.2}&  30.5 & 28.3 &27.3 &25.6& 29.6 & 28.1 &27.2 & 25.6\\
    & & & \scriptsize{0.880} & \scriptsize{0.793} & \scriptsize{0.713} & \scriptsize{0.46} & \scriptsize{0.877} & \scriptsize{0.800} & \scriptsize{0.722} & \scriptsize{0.476} \\\cline{2-11}
    & \multirow{2}{*}{\hdnusplit} &\multirow{2}{*}{11}& 38.4 & 35.9& 34.3& 29.3& 36.8 & 34.9 & 33.8 & 29.3 \\
    & & & \scriptsize{0.981} & \scriptsize{0.966}& \scriptsize{0.951}& \scriptsize{0.844}& \scriptsize{0.975} & \scriptsize{0.962} & \scriptsize{0.948} & \scriptsize{0.843}\\\cline{2-11}
    & \multirow{2}{*}{\usplitKL \textit{\textbf{(ours)}}} &\multirow{2}{*}{1.4}& 38.9 & 35.8 & 35.0 & 30.4 &37.4 & 35.6 & 34.3 & 30.4\\
    & & & \scriptsize{0.985} & \scriptsize{0.968} & \scriptsize{0.960} & \scriptsize{0.867} & \scriptsize{0.979} & \scriptsize{0.968} & \scriptsize{0.953} & \scriptsize{0.865}\\\cline{2-11}
    & \multirow{2}{*}{\denoiSplit \textit{\textbf{(ours)}}} &\multirow{2}{*}{1.6}& \textbf{40.1} & \textbf{37.3} & \textbf{35.7} & \textbf{30.6} & \textbf{38.1} & \textbf{36.6} & \textbf{35.2} & \textbf{30.7}\\
    & & & \textbf{\scriptsize{0.986}} & \textbf{\scriptsize{0.973}} & \textbf{\scriptsize{0.962}} & \textbf{\scriptsize{0.872}} & \textbf{\scriptsize{0.981}} & \textbf{\scriptsize{0.971}} & \textbf{\scriptsize{0.958}} & \textbf{\scriptsize{0.872}}\\\hline
    \multicolumn{6}{c}{} \\[-10pt]
    \hline
    \multirow{8}{*}{T4} &\multirow{2}{*}{\muSplit} &\multirow{2}{*}{7}&  25.9 & 24.3 & 23.6 & 22.4 & 25.2 & 24.2 & 23.5 & 22.4\\
    & & &  \scriptsize{0.777} & \scriptsize{0.664} & \scriptsize{0.556} & \scriptsize{0.331} & \scriptsize{0.729} & \scriptsize{0.640} & \scriptsize{0.554} & \scriptsize{0.321}\\\cline{2-11}
    & \multirow{2}{*}{\hdnusplit} &\multirow{2}{*}{10.7}& 28.8 & 27.9 & 27.4 & 25.8 & 28.2 & 27.6 & 27.2 & 25.7\\ 
    & & & \scriptsize{0.852} & \scriptsize{0.817} & \scriptsize{0.790} & \textbf{\scriptsize{0.725}} & \scriptsize{0.840} & \scriptsize{0.810} & \scriptsize{0.787} & \scriptsize{0.716}\\\cline{2-11}
    & \multirow{2}{*}{\usplitKL \textit{\textbf{(ours)}}} &\multirow{2}{*}{1.3}& 29.4 & 28.5 & 27.5 & 25.9 & \textbf{29.0}  & 27.8 & 27.3 & 25.8\\
    & & & \scriptsize{0.858} & \scriptsize{0.824} & \scriptsize{0.786} & \scriptsize{0.718} & \scriptsize{0.849} &  \scriptsize{0.794} & \scriptsize{0.780} & \scriptsize{0.710}\\ \cline{2-11}
    & \multirow{2}{*}{\denoiSplit \textit{\textbf{(ours)}}} &\multirow{2}{*}{1.5}& \textbf{29.6} & \textbf{28.7} & \textbf{27.6}& \textbf{26.0} & \textbf{29.0} & \textbf{28.5} & \textbf{27.9} & \textbf{26.1}\\
    & & & \textbf{\scriptsize{0.868}} & \textbf{\scriptsize{0.835}} & \textbf{\scriptsize{0.787}} & \textbf{\scriptsize{0.725}} & \textbf{\scriptsize{0.854}} & \textbf{\scriptsize{0.828}} & \textbf{\scriptsize{0.799}} & \textbf{\scriptsize{0.727}}\\
    \hline
    \end{tabular}
    \vspace{1mm}
    \caption{\textbf{Quantitative Results.}
    We show quantitative evaluations for joint denoising and splitting experiments. 
    The four corresponding tasks are abbreviated as T1: ER vs.\ CCPs; T2: ER vs.\ MT; T3: CCPs vs.\ MT, T4: F-actin vs.\ ER. 
    For all experiments, we show the PSNR (sub-row 1) and MS-SSIM (sub-row 2) metrics across $8$ noise levels: Gaussian noise levels of $\sigma\in\{1, 1.5, 2, 4\} $ and Poisson noise levels of $\lambda\in\{0, 1000\}$. 
    The best performance per task and noise level is shown in bold. 
    The third column additionally shows the training time on a single Tesla-V100 GPU (in hours). 
    Not only does \denoiSplit perform best, it does at the same time require considerably less training time.
    }
    \label{tab:overallperf}
\end{table}
}

\newcommand\figTeaser{
\begin{figure}[tbh]
\centering
    \includegraphics[width=.9\textwidth]{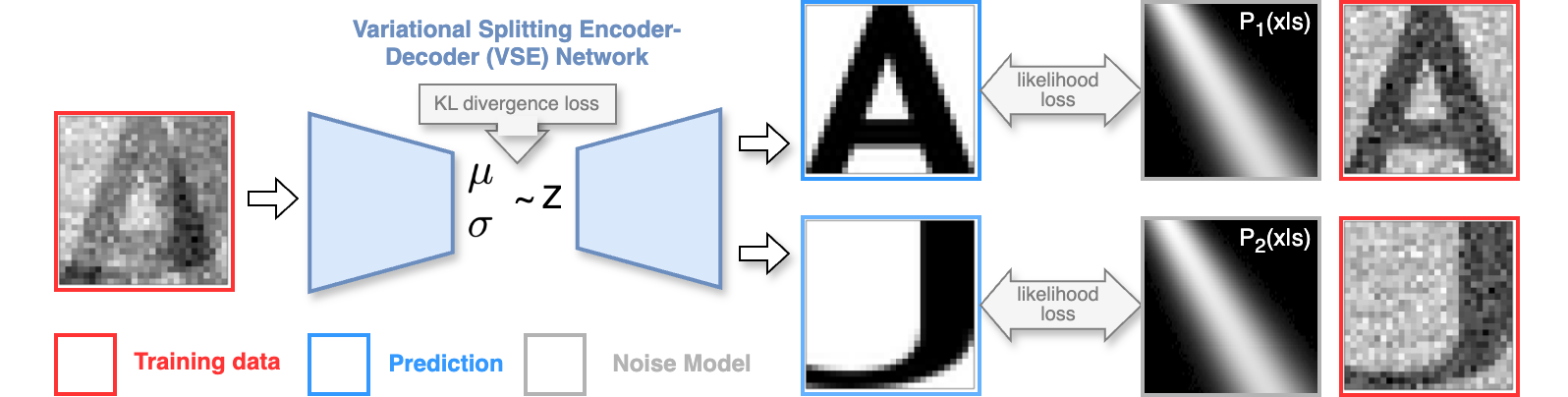}
\caption{\textbf{Teaser Figure.}
In this work we use a variational encoder-decoder network to jointly solve an usupervised denoising and image splitting task and show that our approach outperforms existing baselines.
} 
\label{fig:teaser}
\end{figure}
}

\newcommand\figBioSR{
\begin{figure}[p]
\centering
\begin{overpic}[width=0.8\textwidth]{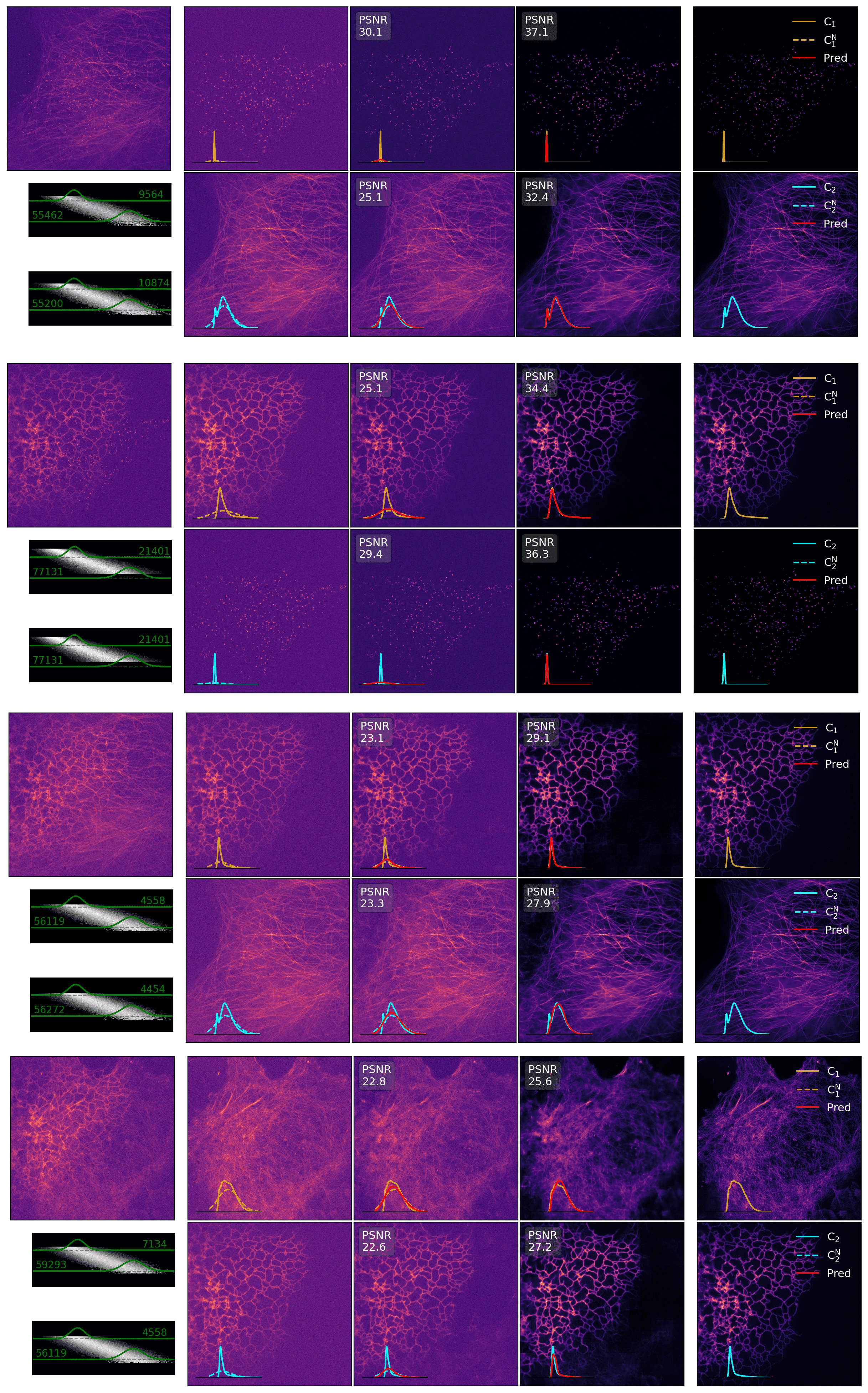}
\put (51,100) {\scriptsize High SNR}
\put (38.3,100) {\scriptsize \denoiSplit}
\put (29, 100){\scriptsize \muSplit}
\put (17.3, 100) {\scriptsize GT}
\put (5, 100) {\scriptsize Input}
\put (-1.6, 97.7) {\scriptsize A}
\put (-1.6, 72.4) {\scriptsize B}
\put (-1.6, 47.1) {\scriptsize C}
\put (-1.6, 21.8) {\scriptsize D}

\put (0.1, 84.3) {\scriptsize $c_0$}
\put (6.3, 81.6) {\scriptsize $c^N_0$}
\put (0, 78.5) {\scriptsize $c_1$}
\put (6.3, 75.3) {\scriptsize $c^N_1$}

\put (0.1, 58.9) {\scriptsize $c_0$}
\put (6.3, 56.0) {\scriptsize $c^N_0$}
\put (0, 52.9) {\scriptsize $c_1$}
\put (6.3, 49.7) {\scriptsize $c^N_1$}

\put (0.1, 33.4) {\scriptsize $c_0$}
\put (6.3, 30.8) {\scriptsize $c^N_0$}
\put (0, 27.3) {\scriptsize $c_1$}
\put (6.3, 24.6) {\scriptsize $c^N_1$}

\put (0.1, 8.7) {\scriptsize $c_0$}
\put (6.3, 6) {\scriptsize $c^N_0$}
\put (0, 2.5) {\scriptsize $c_1$}
\put (6.3, -0.2) {\scriptsize $c^N_1$}

\end{overpic}
\caption{\textbf{Qualitative Results.}
We show examples of noisy inputs, individual noisy channel training data (GT), and predictions by one of the baselines (\muSplit) and our own results obtained with \denoiSplit for four tasks (A: MT vs.\ CCPs, B: ER vs.\ CCPs, C: MT vs.\ ER, and D: F-actin vs.\ ER). 
We show high SNR channel images (not used during training) and show PSNR values \wrt. these images.
Additionally, we plot histograms of various panels for comparison (see legend on the right).
The bottom cell in the first column of each panel shows the used noise models (see main text for details).
The superimposed plots (green) show the distribution of noisy observations ($c_i^N$) for two clean signal intensities.
} 
\label{fig:fullframepredictions}
\end{figure}
}

\newcommand\figSampling{
\begin{figure}[bt]
\centering
\begin{subfigure}{.97\textwidth}
    \begin{overpic}[width=\textwidth]{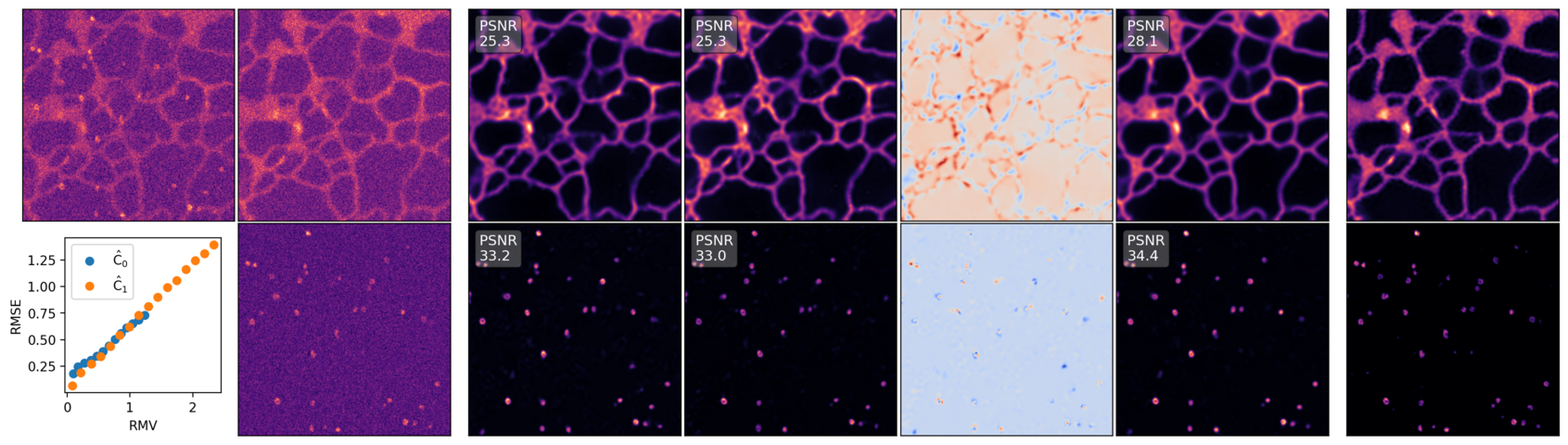}
    \put (87,28.3) {\scriptsize High SNR}
    \put (74.5, 28.3) {\scriptsize MMSE}
    \put (61, 28.3) {\scriptsize $S_1$ - $S_2$}
    \put (45.3, 28.3) {\scriptsize Sample 2}
    \put (32.5, 28.3) {\scriptsize Sample 1}
    \put (20, 28.3) {\scriptsize GT}
    \put (6, 28.3) {\scriptsize Input}
    \put (-1.2, 26.5) {\scriptsize A} 
    \end{overpic}    
\end{subfigure}
\begin{subfigure}{.97\textwidth}
\begin{overpic}[width=\textwidth]{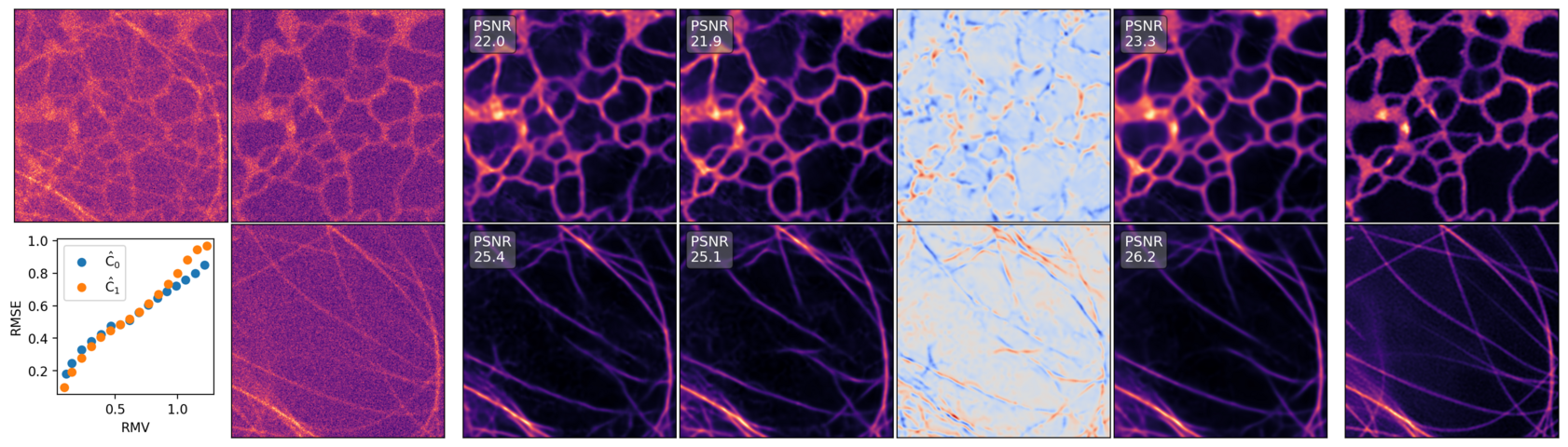}
\put (-1.2, 26.5) {\scriptsize B}
\end{overpic}
\end{subfigure}
\begin{subfigure}{.97\textwidth}
\begin{overpic}[width=\textwidth]{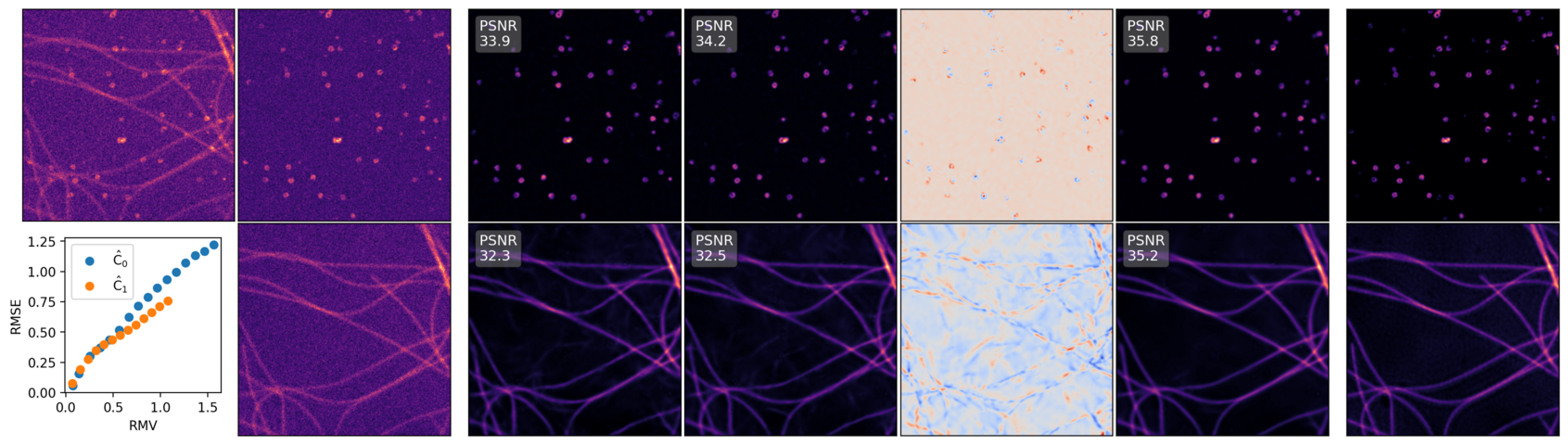}
\put (-1.2, 26.5) {\scriptsize C}
\end{overpic}
\end{subfigure}
\caption{\textbf{Variational Sampling and Calibration.}
The VSE Network in \denoiSplit is capable of sampling from a learned posterior.
Here we show cropped inputs  ($256\times256$), two corresponding prediction samples, the difference between the two samples ($S_1 - S_2$), the MMSE prediction, and otherwise unused high SNR microscopy for three tasks, namely ER vs.\ CCPs, ER vs.\ MT, and CCPs vs.\ MT. 
The MMSE predictions are computed by averaging $50$ samples. 
As before, we show PSNR \wrt high SNR patches. 
The dot plots in the first column show are calibration plots, showcasing that the error estimate we propose works well (see main text).
}
\label{fig:sampling}
\end{figure}
}

\newcommand\figPatchComparison{
\begin{figure}[tbp]
\centering
\begin{overpic}[width=\textwidth]{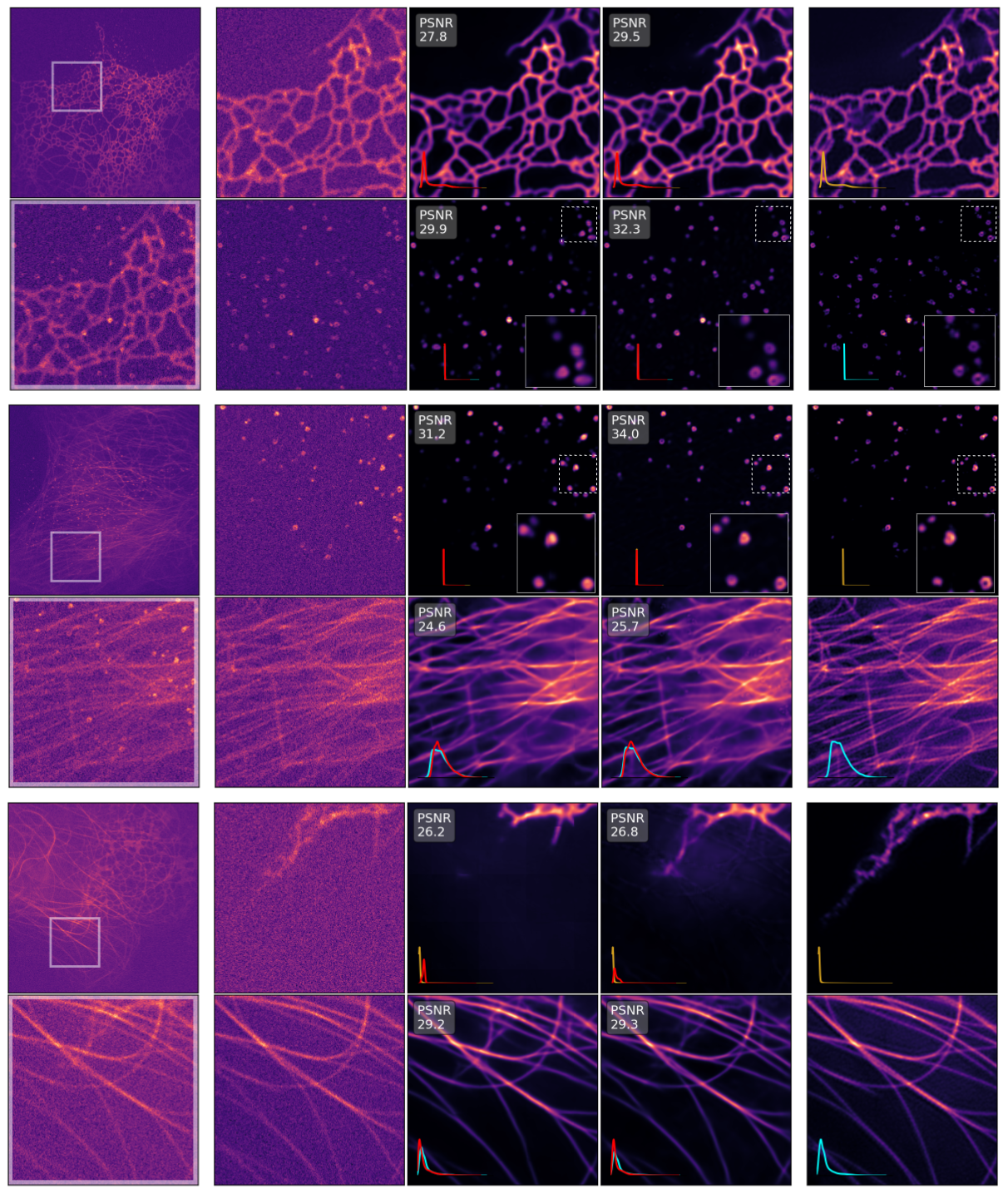} 
\put (70.5,100) {\scriptsize High SNR}
\put (53.7, 100) {\scriptsize \denoiSplit}
\put (36.3, 100) {\scriptsize \hdnusplit}
\put (24, 100) {\scriptsize GT}
\put (6, 100) {\scriptsize Input}
\put (-1.2, 97.5) {\scriptsize A}
\put (-1.2, 64) {\scriptsize B}
\put (-1.2, 30.7) {\scriptsize C}
\end{overpic}
\caption{\textbf{Comparison to Sequential Baseline.}
For each panel (ER vs.\ CCPs, CCPs vs.\ MT, and ER vs.\ MT) we show the full input image and its ($256\times256$) inset crop, corresponding noisy training data crops (GT), the results of the sequential denoising and splitting baseline (\hdnusplit) and our end-to-end results obtained with \denoiSplit. 
All predictions show the MMSE, obtained by averaging $50$ sampled predictions. 
We show a few zoomed-in locations where the baseline under-performs.
Note that such small differences might contribute little to evaluations via PSNR, but can make a huge difference for the downstream analysis of investigated biological structures contained in such microscopy data.
}
\label{fig:patchcomparison}
\end{figure}
}
\newcommand\realNoisePerformance{
\begin{figure}[h]
\begin{minipage}{.65\textwidth}
\begin{overpic}[width=\textwidth]{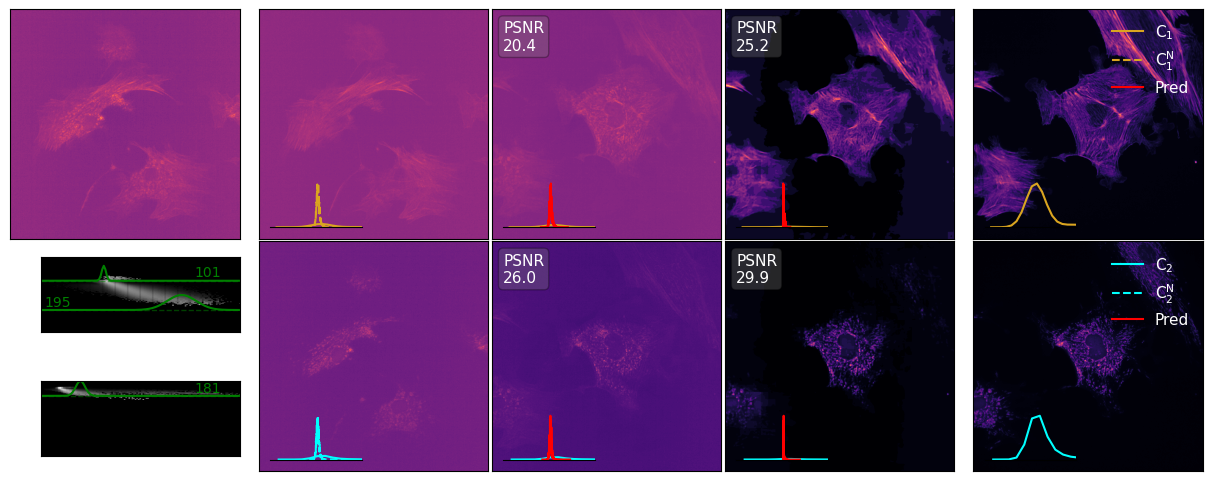}  
\put (81,40) {\scriptsize High SNR}
\put (63,40) {\scriptsize \denoiSplit}
\put (46, 40){\scriptsize \muSplit}
\put (29, 40) {\scriptsize GT}
\put (7, 40) {\scriptsize Input}
\end{overpic}
\end{minipage}%
\hspace{2mm}
\begin{minipage}{.3\textwidth}
    \begin{tabular}{c|c}
        Model & Eval. \\
        \hline
        \multirow{2}{*}{\muSplit} & 26.5 \\
        & \scriptsize{0.872} \\
        \hline
         \multirow{2}{*}{\hdnusplit}  & 28.1 \\
         & \scriptsize{0.887} \\
         \hline
         \multirow{2}{*}{Altered \muSplit} & 31.1\\
         & \scriptsize{0.936} \\
         \hline
         \multirow{2}{*}{\denoiSplit} & 31.0 \\
         & \scriptsize{0.935} \\
         
    \end{tabular}
\end{minipage}
    \caption{\textbf{Results on Actin \vs Mito Task:} \textbf{(Left)} Here, qualitative evaluation of the different models on Hagen et al.~\cite{Hagen2021-xh} is shown. We also show High SNR channel images (not used during training) in last column and we show PSNR \wrt. them. Noise models are shown in column one, second row. \textbf{(Right)} Quantitative evaluation of \denoiSplit along with the baselines using PSNR (line 1) and range invariant MS-SSIM~\cite{Weigert2018-pi} (line 2, also see \supsection{2} for details on the MS-SSIM variant). Note that \HDN training in \hdnusplit was quite unstable and so,  we had to train it with a lower hierarchy count (3 as opposed to default 6).}
\label{fig:realnoise}
\end{figure}
}
\newcommand\tabOverallPerfOtherDset{
\begin{table}[]
    \centering
    \begin{tabular}{|c|l|c|c|c|c|c|c|c|c|c|}
    \hline
    \multirow{3}{*}{Task}&\multirow{3}{*}{Model} & \multirow{3}{*}{training} & \multicolumn{8}{c|}{Noise level parameters}\\\cline{4-11}   
    & & & \multicolumn{4}{c|}{$\lambda=0$} & \multicolumn{4}{c|}{$\lambda=1000$}\\\cline{4-11}      
    & & $[h]$ &$\sigma=1$ & 1.5 & 2 & 4 & $\sigma=1$ & 1.5 & 2 & 4\\
    \hline
    \multicolumn{6}{c}{} \\[-10pt]
    \hline
    \multirow{8}{*}{T5} &\multirow{2}{*}{\muSplit} &\multirow{2}{*}{5.5} & 25.1 & 23.7 & 23.1 & 22.1 & 24.8 & 23.8 & 23.0 & 22.1\\\cline{4-11}
    &  & & \scriptsize{0.728} & \scriptsize{0.633} & \scriptsize{0.537} & \scriptsize{0.341} & \scriptsize{0.697} & \scriptsize{0.593} & \scriptsize{0.536} & \scriptsize{0.307}\\ \cline{2-11}
    & \multirow{2}{*}{\hdnusplit} &\multirow{2}{*}{8} & 27.3 & 26.6 & \textbf{26.1} & \textbf{24.6} & 27.4 & 26.4 & \textbf{26.1} & 24.3\\
    &  & & \scriptsize{0.793} & \scriptsize{0.756} & \textbf{\scriptsize{0.731}} & \textbf{\scriptsize{0.645}} & \textbf{\scriptsize{0.794}} & \scriptsize{0.739} & \textbf{\scriptsize{0.722}} & \textbf{\scriptsize{0.639}} \\\cline{2-11}
    & \multirow{2}{*}{\usplitKL \textit{\textbf{(ours)}}} &\multirow{2}{*}{1.3}& \textbf{27.9} & 26.8 & 25.9 & 24.5 & 27.4 & 26.5 & 25.9 & \textbf{24.4}  \\\cline{4-11}
    & & & \textbf{\scriptsize{0.799}} & \scriptsize{0.741} & \scriptsize{0.698} & \scriptsize{0.601} & \scriptsize{0.780} & \scriptsize{0.731} & \scriptsize{0.697} & \scriptsize{0.595} \\\cline{2-11}
    &\multirow{2}{*}{\denoiSplit \textit{\textbf{(ours)}}} &\multirow{2}{*}{1.5}& 27.7 & \textbf{27.0} & 25.9 & 24.5 & \textbf{27.5} & \textbf{27.1} & 26.0 & \textbf{24.4} \\\cline{4-11}
    &  & & \scriptsize{0.798} & \textbf{\scriptsize{0.762}} & \scriptsize{0.707} & \scriptsize{0.604} & \scriptsize{0.787} & \textbf{\scriptsize{0.761}} & \scriptsize{0.710} & \scriptsize{0.606}\\\hline
    \hline
    \multirow{8}{*}{T6} &\multirow{2}{*}{\muSplit} &\multirow{2}{*}{5} & 30.5 & 28.7 & 27.8 & 26.7 & 29.9 & 28.7 & 27.9 & 26.8\\\cline{4-11}
    &  & & \scriptsize{0.869} & \scriptsize{0.779} & \scriptsize{0.685} & \scriptsize{0.422} & \scriptsize{0.866} & \scriptsize{0.788} & \scriptsize{0.708} & \scriptsize{0.443}\\ \cline{2-11}
    & \multirow{2}{*}{\hdnusplit} &\multirow{2}{*}{7} & 34.6 & 33.1 & 32.1 & 28.8 & 33.8 & 32.6 & 32.0 & 28.8\\
    &  & & \scriptsize{0.938} & \scriptsize{0.907} & \scriptsize{0.885} & \scriptsize{0.785} & \scriptsize{0.926} &  \scriptsize{0.899} & \scriptsize{0.882} & \scriptsize{0.783}\\\cline{2-11}
    & \multirow{2}{*}{\usplitKL \textit{\textbf{(ours)}}} &\multirow{2}{*}{1.3}& 35.5 & 33.3 & 32.5 & \textbf{30.4} & 34.4 & 32.8 & 32.4 & \textbf{30.1} \\\cline{4-11}
    & & & \textbf{\scriptsize{0.950}} & \scriptsize{0.911} & \scriptsize{0.890} & \scriptsize{0.825} & \scriptsize{0.935} & \scriptsize{0.898} & \scriptsize{0.887} & \scriptsize{0.819}\\\cline{2-11}
    &\multirow{2}{*}{\denoiSplit \textit{\textbf{(ours)}}} &\multirow{2}{*}{1.5}&\textbf{36.3} & \textbf{34.4} & \textbf{33.1} & \textbf{30.4} & \textbf{34.7} & \textbf{33.8} & \textbf{32.9} & 29.9 \\\cline{4-11}
    &  & & \scriptsize{0.945} & \textbf{\scriptsize{0.927}} & \textbf{\scriptsize{0.895}} & \textbf{\scriptsize{0.830}} & \textbf{\scriptsize{0.939}} & \textbf{\scriptsize{0.918}} & \textbf{\scriptsize{0.890}} & \textbf{\scriptsize{0.826}} \\\hline
    \end{tabular}
    \vspace{1mm}
    \caption{\textbf{Quantitative Results.}
    We show quantitative evaluations for two more tasks which are abbreviated as T5: F-actin vs.\ MT and T6: F-actin vs CCPs. 
    For all experiments, we show the PSNR (sub-row 1) and MS-SSIM (sub-row 2) metrics across $8$ noise levels: Gaussian noise levels of $\sigma\in\{1, 1.5, 2, 4\} $ and Poisson noise levels of $\lambda\in\{0, 1000\}$. 
    The best performance per task and noise level is shown in bold. 
    The third column additionally shows the training time on a single Tesla-V100 GPU (in hours). 
    }
    \label{tab:overallperfOtherdset}
\end{table}
}

\newcommand\tabMoreTasks{
\begin{table}[]
    \centering
    \begin{tabular}{l|c|c|c|c}
    Task & $\sigma=1$ & $\sigma=1.5$ & $\sigma=2$ & $\sigma=4$ \\
    \hline
    ER\vs CCPs &  3400 & 5100 & 6800 & 13600  \\
    ER \vs MT & 4450 & 6675 & 8900 & 17800   \\
    CCPs \vs MT & 3150 & 4725 & 6300 & 12600\\
    F-actin \vs ER &4450 & 6675 & 8900 & 17800 \\
    F-actin \vs CCPs & 3050 & 4575 & 6100 & 12200 \\
    F-actin \vs MT & 4300 & 6450 & 8600 & 17200 \\
    \end{tabular}
    \caption{Gaussian $\sigma$ values for the different tasks. Note that they have been estimated by computing the standard deviation on the input images of these tasks.}
    \label{tab:gaussianSigmaValues}
\end{table}
}

\newcommand\tabOtherDsets{
\begin{table}[]
    \centering
    \begin{tabular}{c|c|c|c|c|c|c|c|c}
    \multirow{3}{*}{Model} & \multicolumn{8}{c}{Tasks} \\\cline{2-9}
     & \multicolumn{4}{c|}{T7} & \multicolumn{4}{c|}{T8} \\\cline{2-9}
     &  1 & 1.5 & 2 & 4 & 1 & 1.5 & 2 & 4 \\
    \hline
    \multicolumn{9}{c}{} \\[-10pt]
    \hline

     \multirow{2}{*}{\muSplit} & 22.6 & 21.1& 20.2& 19.1&  28.7& 27.0 & 26.2 & 25.2\\
     &  \scriptsize{0.555} & \scriptsize{0.442} & \scriptsize{0.361} & \scriptsize{0.189} & \scriptsize{0.905} & \scriptsize{0.825} & \scriptsize{0.747} & \scriptsize{0.489} \\
     \hline
     \multirow{2}{*}{\hdnusplit} & 27.8 & 27.3 & 27.0& 26.4 & - & - & - & - \\
      & \scriptsize{0.880}& \scriptsize{0.871}& \scriptsize{0.865} & \scriptsize{0.843} &-&-&-&-\\
      \hline
     \multirow{2}{*}{(Ours) \usplitKL} & 26.3 & 26.2 & 25.9 & 25.2 & 34.7 & 34.2 & 33.6 & 32.3 \\
      & \scriptsize{0.838} & \scriptsize{0.826} & \scriptsize{0.820} & \scriptsize{0.807} & \scriptsize{0.975} & \scriptsize{0.971} & \scriptsize{0.966} & \scriptsize{0.951} \\
      \hline
    \multirow{2}{*}{(Ours) denoisplit} &26.4 & 26.1 & 26.0 & 25.2 & 35.5 & 33.7 & 33.5 & 32.2\\  
     &  \scriptsize{0.835} & \scriptsize{0.827} & \scriptsize{0.825} & \scriptsize{0.807} & \scriptsize{0.979} & \scriptsize{0.974} & \scriptsize{0.968} & \scriptsize{0.951} \\  
    \hline

    \end{tabular}
\caption{T7: Actin vs Tubulin from PaviaATN dataset, T8: Actin vs Mito High-SNR. For T8, \HDN training was quite unstable and crashed multiple times due to NaNs. Due to this reason, there are no entries for \hdnusplit for the task T8. Note that Actin vs Mito los-SNR task, which is present in the main text, also had trouble training \HDN. }
\label{tab:overallperfothertable}
\end{table}
}

\newcommand\tableDeraining{
\begin{table}[h]
\centering
\begin{tabular}{c|c|c|c|c|c}
& ID~\cite{derain-ID} & LP~\cite{derain-Li_2016_CVPR} & DSC~\cite{derain-dsc} & JORDER-R~\cite{derain-jorder} & \denoiSplit \\
\hline
PSNR &14.02 & 14.26 & 15.66 & 23.45 & 26.2\\
SSIM & 0.5239 & 0.4225 & 0.5444 & 0.7490 & 0.758\\
\end{tabular}
\caption{Quantitative results on Rain100H dataset~\cite{derain-jorder} for De-raining task. Due to the absence of noise, we disabled the noise model in our \denoiSplit. Metric values of all other methods have been taken from~\cite{derain-jorder}.}
\label{tab:deraining}
\end{table}
}

\newcommand\tableHazing{
\begin{table}[h]
\centering
\begin{tabular}{c|c|c|c|c|c}
& DM$^2$F-Net~\cite{dehazedm2fnet} & FFA-Net~\cite{dehazeQin2019FFANetFF} & DA~\cite{dehazingDA} & DMT-Net~\cite{dehazing-dmtnet} & \denoiSplit \\
\hline
PSNR &24.61 & 26.97 & 24.03 & 28.53 & 27.1\\
SSIM &0.92 & 0.95 & 0.90 & 0.96 & 0.90\\
\end{tabular}
\caption{Quantitative results on Haze4K dataset~\cite{dehazing-dmtnet} for De-hazing task. Due to the absence of noise, we disabled the noise model in our \denoiSplit. Metric values of all other methods have been taken from~\cite{dehazing-dmtnet}.}
\label{tab:dehazing}
\end{table}
}
\newcommand\figDenoisingSplittingComparison{
\begin{figure}[tbh]
\centering
    \includegraphics[width=.9\textwidth]{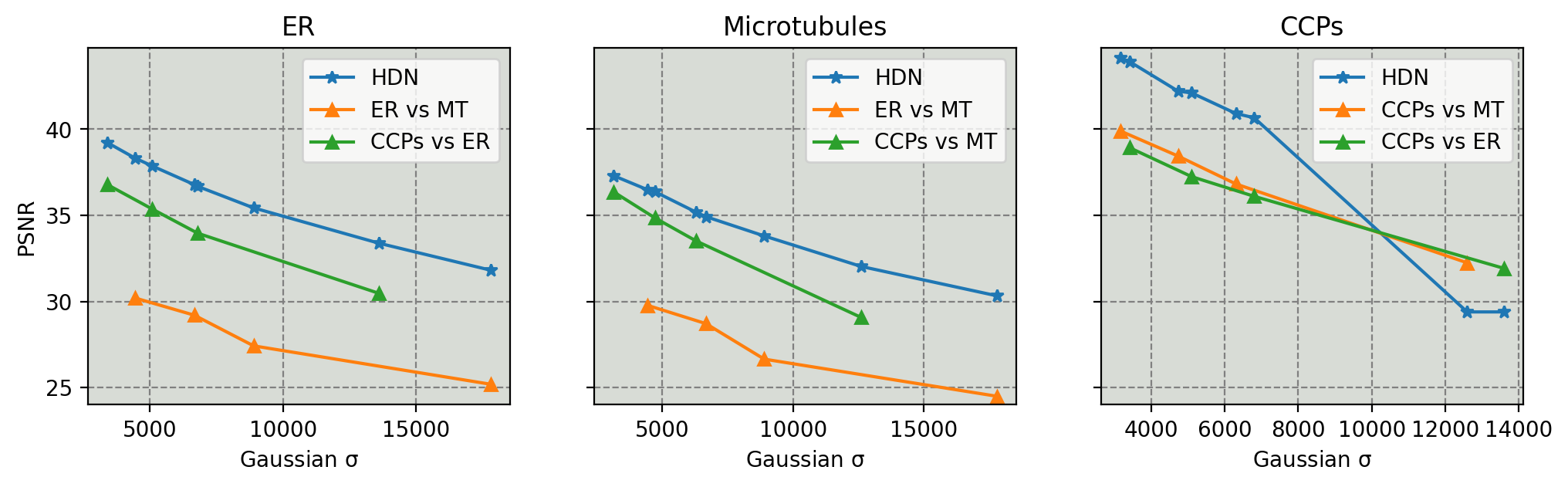}
\caption{\textbf{Problem comparison: Joint Denoising-Splitting \vs.\ Denoising}
In this figure, we compare the PSNR with respect to high-SNR micrographs for two image restoration tasks: (a) Self-supervised denoising for which we use \HDN and (b) joint denoising-splitting where \denoiSplit is used. We evaluate the two models over multiple noise levels of Gaussian noise (x-axis) with Poisson($\lambda=1000$) noise present in all cases. For \denoiSplit, we use three tasks namely: ER \vs.\ MT, CCPs \vs ER and CCPs vs MT. We show one plot for each structure type (ER, Microtubules and CCPs). Two things are evident: (a) Judging just from PSNR numbers, we can say that Denoising is a simpler task than joint Denoising-Splitting. (b) Performance of \denoiSplit for one channel depends on the other channel as well. For example, for \denoiSplit, PSNR on ER channel (first plot) is higher when the task is CCPs \vs ER (green) as opposed to ER \vs MT (orange) task. 
} 
\label{fig:denoisingVSsplitting}
\end{figure}
}

\newcommand\figFactinER{
\begin{figure}[h]
\begin{overpic}[width=\textwidth]{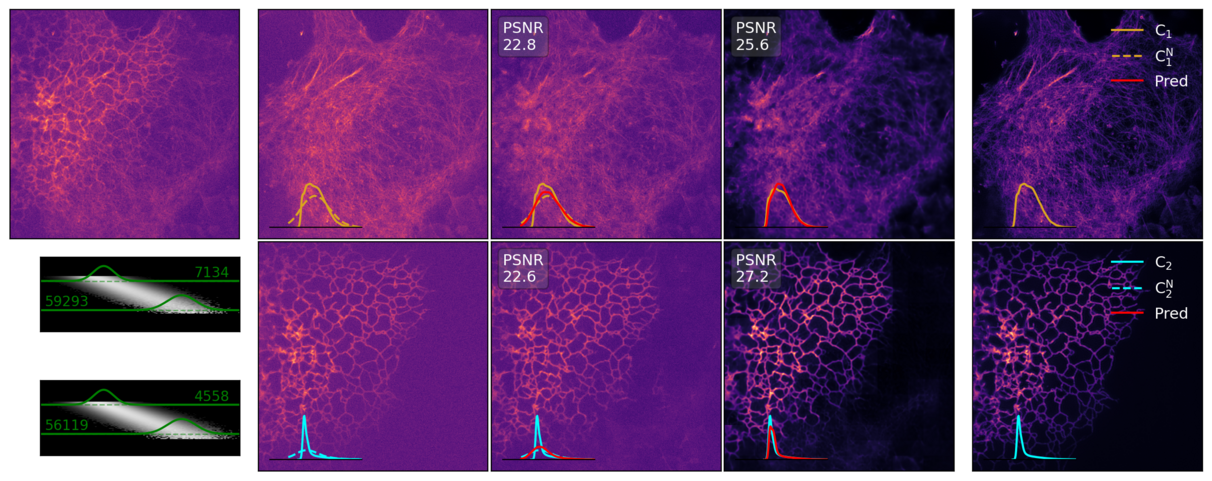} 
\put (86,40) {\scriptsize High SNR}
\put (66,40) {\scriptsize \denoiSplit}
\put (48, 40){\scriptsize \muSplit}
\put (30, 40) {\scriptsize GT}
\put (8, 40) {\scriptsize Input}
\end{overpic}
\caption{\textbf{Qualitative Results F-actin \vs ER:} In this figure, we show full frame prediction on F-actin \vs ER task. We show noisy input (column one), individual noisy channel training data (column two), and predictions by one of the baselines \muSplit (column three) and our own results obtained with \denoiSplit (column four). 
Additionally we show high SNR channel images (not used during training) as the last column and show PSNR values \wrt. these images.
Additionally, we plot histograms of pixel intensities various panels for comparison (see legend on the right).
The second row, first column shows the used noise models.
The superimposed plots (green) show the distribution of noisy observations ($c_i^N$) for two clean signal intensities. }
\label{fig:factinVsER}
\end{figure}
}

\newcommand\figFactinMT{
\begin{figure}[h]
\begin{overpic}[width=\textwidth]{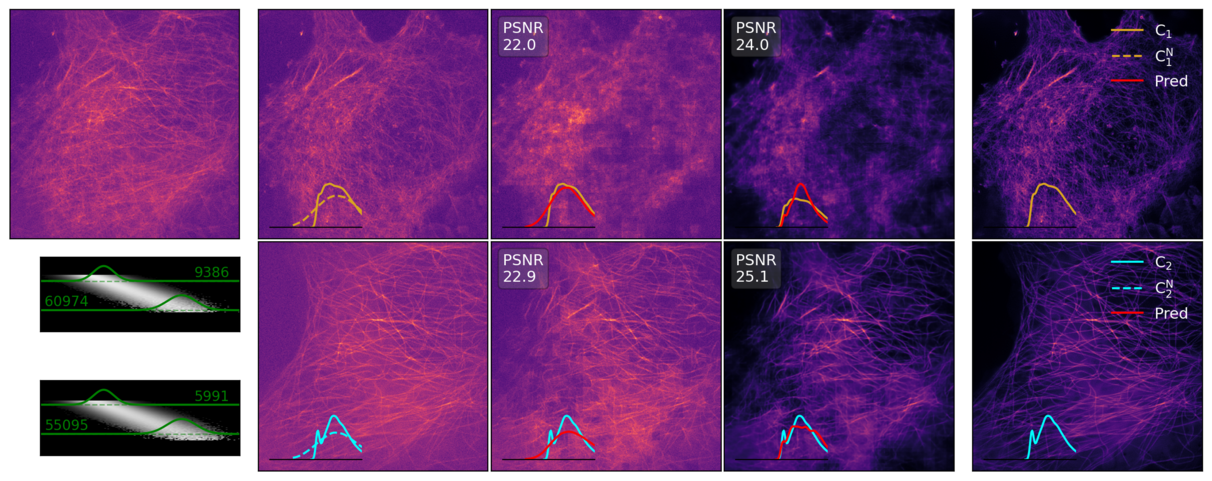}
\put (86,40) {\scriptsize High SNR}
\put (66,40) {\scriptsize \denoiSplit}
\put (48, 40){\scriptsize \muSplit}
\put (30, 40) {\scriptsize GT}
\put (8, 40) {\scriptsize Input}

\end{overpic}

\caption{\textbf{Qualitative Results F-actin \vs MT:} In this figure, we show full frame prediction on F-actin \vs MT task. We show noisy input (column one), individual noisy channel training data (column two), and predictions by one of the baselines \muSplit (column three) and our own results obtained with \denoiSplit (column four). 
Additionally we show high SNR channel images (not used during training) as the last column and show PSNR values \wrt. these images.
Additionally, we plot histograms of pixel intensities various panels for comparison (see legend on the right).
The second row, first column shows the used noise models.
The superimposed plots (green) show the distribution of noisy observations ($c_i^N$) for two clean signal intensities. }
\label{fig:factinVsMT}
\end{figure}
}

\newcommand\figFactinCCPs{
\begin{figure}[h]
\begin{overpic}[width=\textwidth]{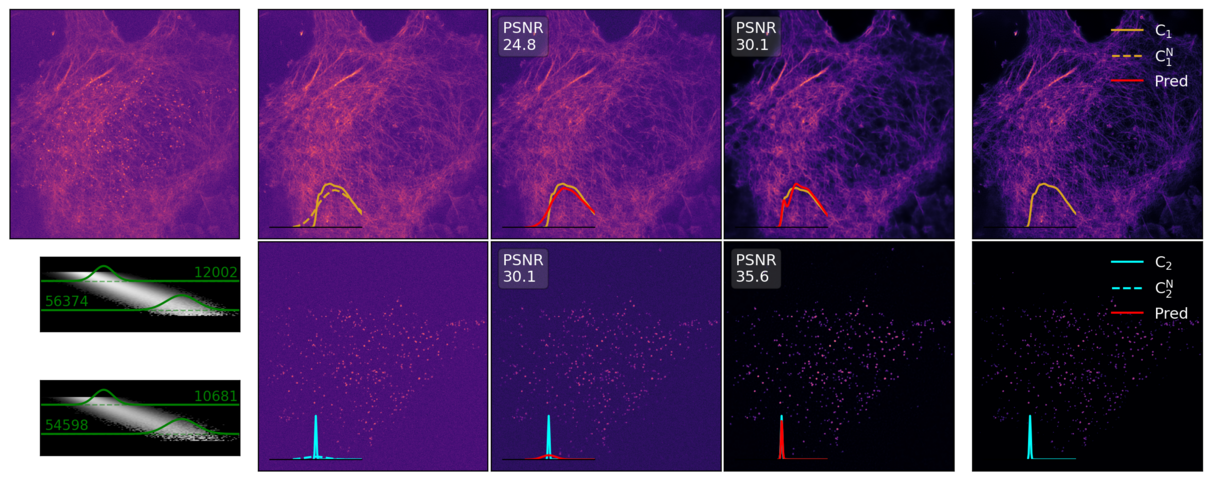} 
\put (86,40) {\scriptsize High SNR}
\put (66,40) {\scriptsize \denoiSplit}
\put (48, 40){\scriptsize \muSplit}
\put (30, 40) {\scriptsize GT}
\put (8, 40) {\scriptsize Input}
\end{overpic}

\caption{\textbf{Qualitative Results F-actin \vs CCPs:} In this figure, we show full frame prediction on F-actin \vs CCPs task. We show noisy input (column one), individual noisy channel training data (column two), and predictions by one of the baselines \muSplit (column three) and our own results obtained with \denoiSplit (column four). 
Additionally we show high SNR channel images (not used during training) as the last column and show PSNR values \wrt. these images.
Additionally, we plot histograms of pixel intensities various panels for comparison (see legend on the right).
The second row, first column shows the used noise models.
The superimposed plots (green) show the distribution of noisy observations ($c_i^N$) for two clean signal intensities. }
\label{fig:factinVsCCPs}
\end{figure}
}

\newcommand\figrealNoise{
\begin{figure}[h]
\begin{overpic}[width=\textwidth]{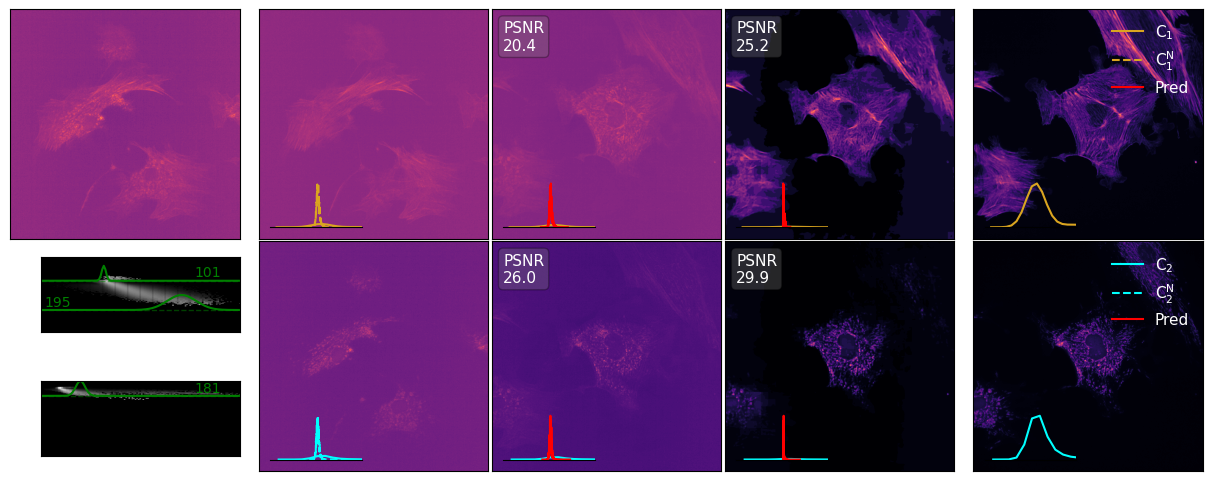}  
\put (86,40) {\scriptsize High SNR}
\put (66,40) {\scriptsize \denoiSplit}
\put (48, 40){\scriptsize \muSplit}
\put (30, 40) {\scriptsize GT}
\put (8, 40) {\scriptsize Input}
\end{overpic}

\caption{\textbf{Qualitative Results Actin \vs Mito:} In this figure, we show full frame prediction on Actin \vs Mitochondria task. Here, the noise in the target channels is not synthetic but is real microscopy noise. We show noisy input (column one), individual noisy channel training data (column two), and predictions by one of the baselines \muSplit (column three) and our own results obtained with \denoiSplit (column four). 
Additionally we show high SNR channel images (not used during training) as the last column and show PSNR values \wrt. these images.
Additionally, we plot histograms of pixel intensities various panels for comparison (see legend on the right).
The second row, first column shows the used noise models.
The superimposed plots (green) show the distribution of noisy observations ($c_i^N$) for two clean signal intensities. }
\label{fig:realnoise}
\end{figure}
}

\newcommand\figCalibrationMmseVariation{
\begin{figure}[h]
\centering
\includegraphics[width=0.75\textwidth]{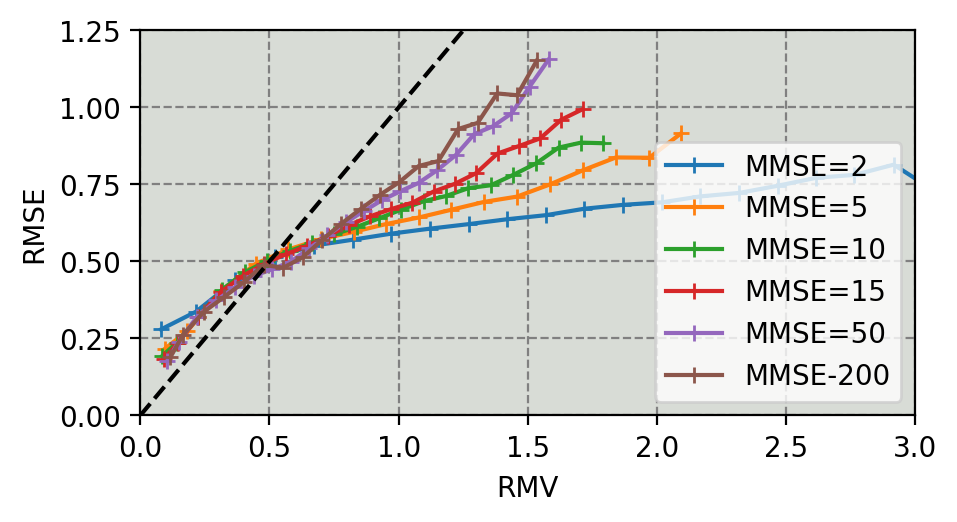} 
\caption{In this figure, we show that as we increase the sample count to get the uncalibrated estimate of pixelwise uncertainty, the calibration diagram as shown in this plot becomes better and better, \ie, gets more and more closer to $y=x$.}
\label{fig:calibmmse}
\end{figure}
}

\newcommand\figNoiseModelComparison{
\begin{figure}[!tbp]
  \centering
  \begin{minipage}[b]{0.45\textwidth}
    \includegraphics[width=\textwidth]{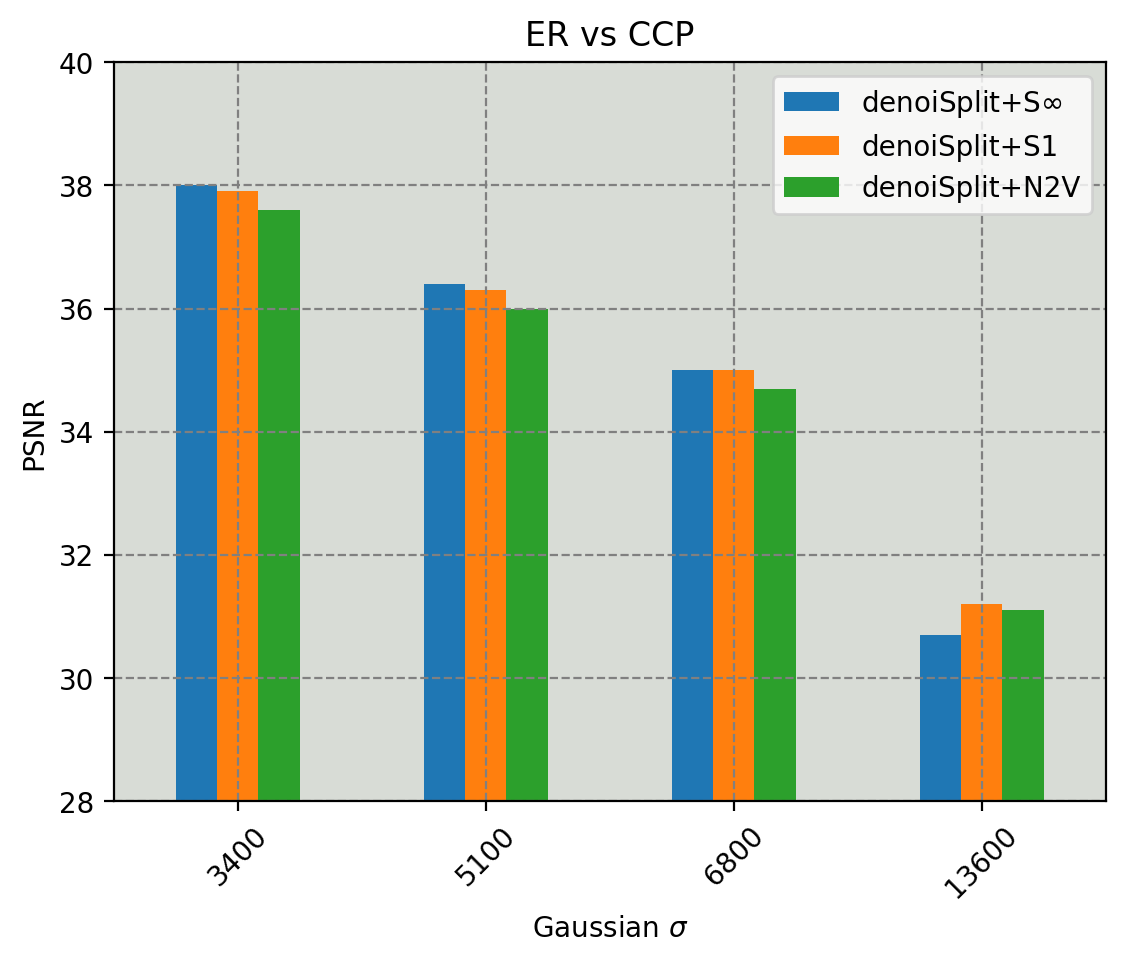}
  \end{minipage}
  \hfill
  \begin{minipage}[b]{0.45\textwidth}
    \includegraphics[width=\textwidth]{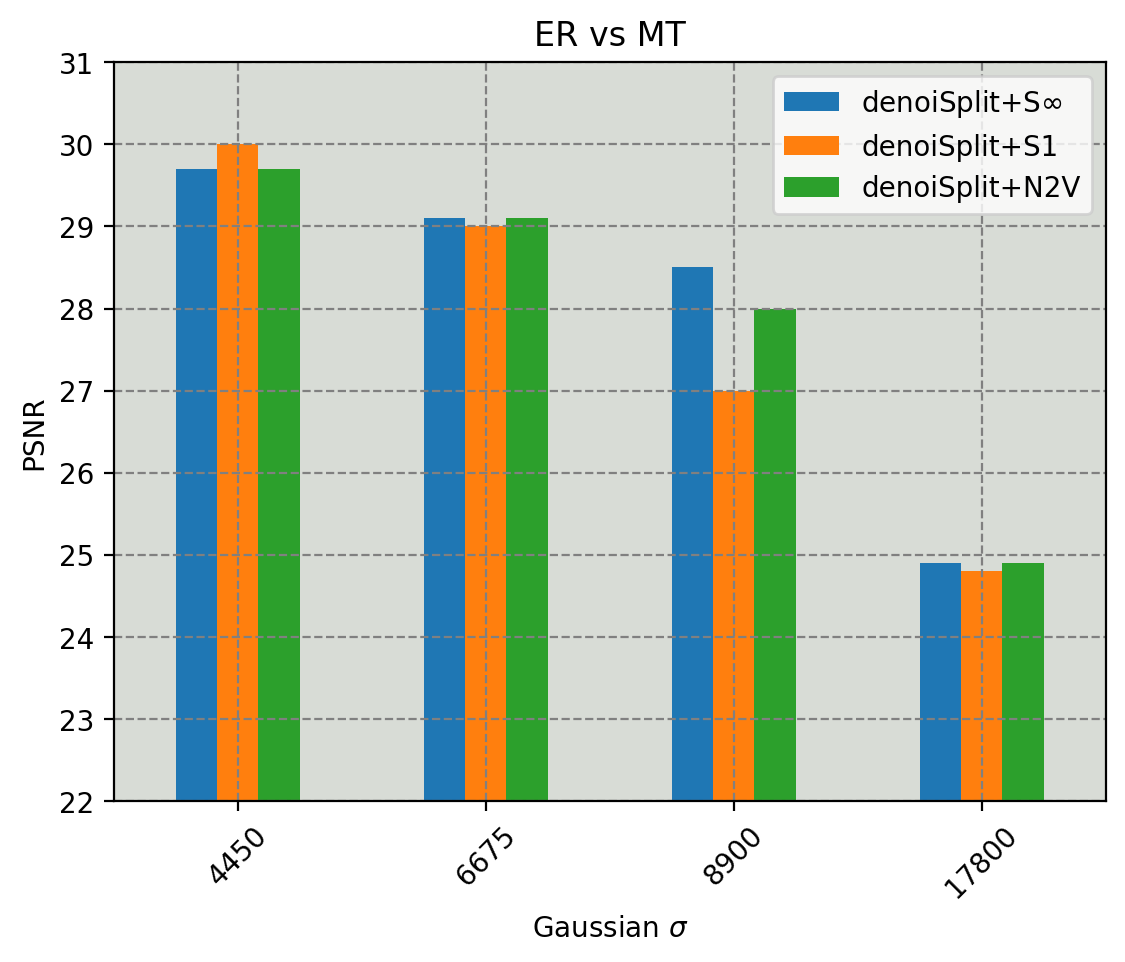}
  \end{minipage}
\caption{\textbf{Quantitative comparison of different noise model generation methodologies}: Here, we compare three ways in which noise models can be generated. (a) denoiSplit$+S_\infty$: When for every clean pixel intensity, one has access to multiple noisy intensities. This corresponds to the case when one has access to the microscope which has generated the data. (b) denoiSplit$+S1$: This corresponds the case when one has access to clean data and its corresponding noisy data. (c) denoiSplit$+N2V$: This corresponds to the case when one has access to just noisy data. We use N2V to denoise them which we use as clean data. We compare \denoiSplit trained using each of the three noise model variants. We show PSNR performance on two tasks.}
\label{fig:noiseModelMultipleWays}
\end{figure}
}

\newcommand\figHDNERvsCCPs{
\begin{figure}[!tbp]
  \centering
  \begin{minipage}[b]{0.32\textwidth}
    \includegraphics[width=\textwidth]{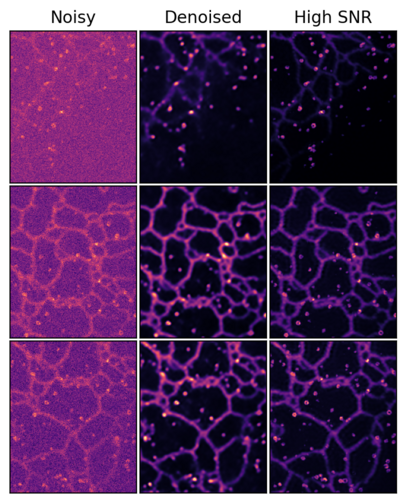}
  \end{minipage}
  \hfill
  \begin{minipage}[b]{0.32\textwidth}
    \includegraphics[width=\textwidth]{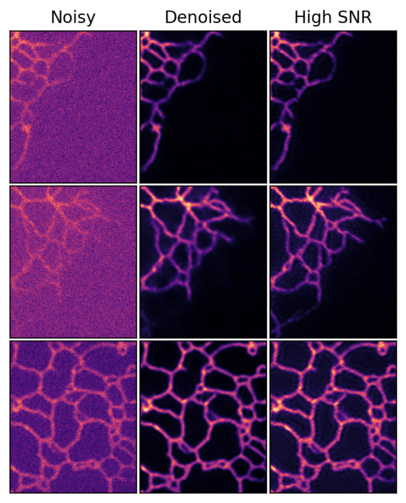}
  \end{minipage}
  \hfill
  \begin{minipage}[b]{0.32\textwidth}
    \includegraphics[width=\textwidth]{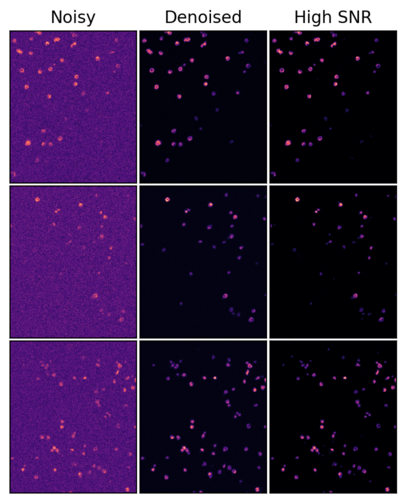}
  \end{minipage}
  
\caption{Qualitative performance of HDN on ER \vs CCPs task input and its two constituent channels. Left panel shows the denoising performance on input for our splitting task and central and right panel shows its denoising performance on its two constituent channels. Within each panel, we show three random patches (rows) of size $256\times213$. Specifically, we show the input (first column), denoised predictions (second column) and high SNR patch (last column).}
\label{fig:HDN_perf_ERvsCCPs}
\end{figure}
}

\newcommand\figHDNERvsMT{
\begin{figure}[!tbp]
  \centering
  \begin{minipage}[b]{0.32\textwidth}
    \includegraphics[width=\textwidth]{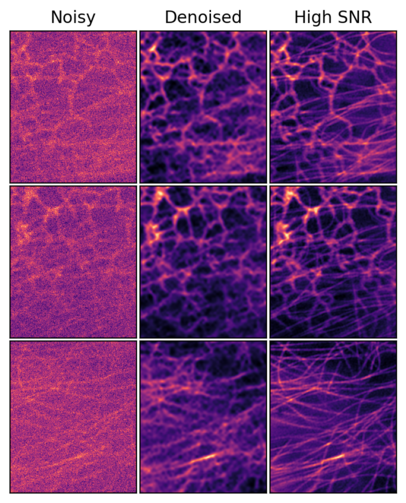}
  \end{minipage}
  \hfill
  \begin{minipage}[b]{0.32\textwidth}
    \includegraphics[width=\textwidth]{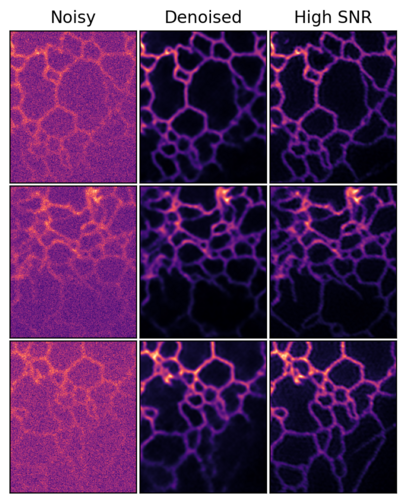}
  \end{minipage}
  \hfill
  \begin{minipage}[b]{0.32\textwidth}
    \includegraphics[width=\textwidth]{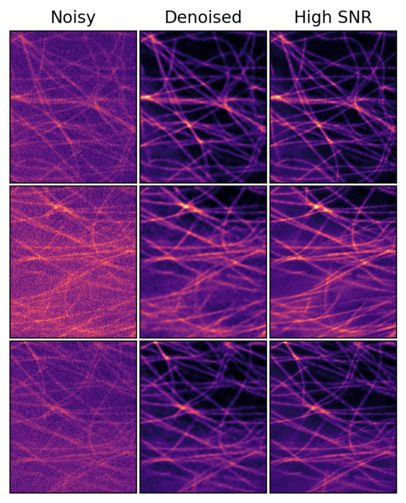}
  \end{minipage}
  
\caption{Qualitative performance of HDN on ER \vs MT task input and its two constituent channels. Left panel shows the denoising performance on input for our splitting task and central and right panel shows its denoising performance on its two constituent channels. Within each panel, we show three random patches (rows) of size $256\times213$. Specifically, we show the input (first column), denoised predictions (second column) and high SNR patch (last column).}
\label{fig:HDN_perf_ERvsMT}
\end{figure}
}

\newcommand\figHDNCCPsvsMT{
\begin{figure}[!tbp]
  \centering
  \begin{minipage}[b]{0.32\textwidth}
    \includegraphics[width=\textwidth]{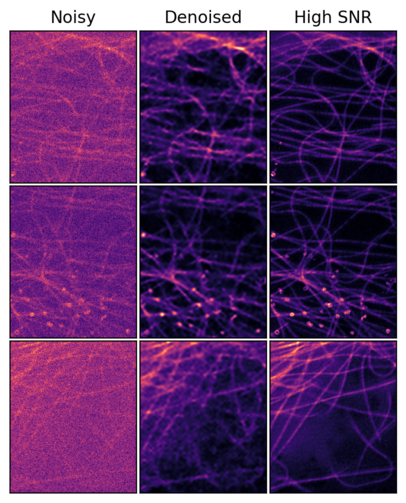}
  \end{minipage}
  \hfill
  \begin{minipage}[b]{0.32\textwidth}
    \includegraphics[width=\textwidth]{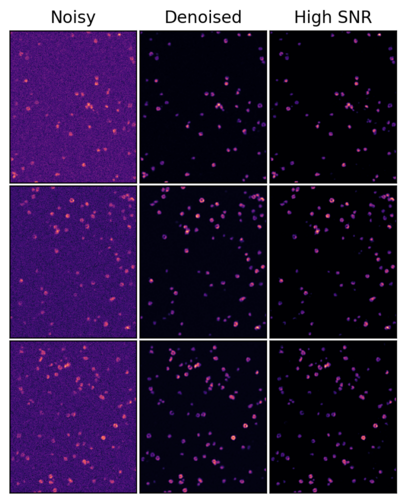}
  \end{minipage}
  \hfill
  \begin{minipage}[b]{0.32\textwidth}
    \includegraphics[width=\textwidth]{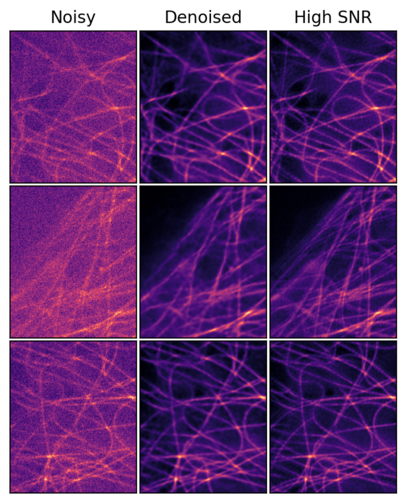}
  \end{minipage}
  
\caption{Qualitative performance of HDN on CCPs \vs MT task input and its two constituent channels. Left panel shows the denoising performance on input for our splitting task and central and right panel shows its denoising performance on its two constituent channels. Within each panel, we show three random patches (rows) of size $256\times213$. Specifically, we show the input (first column), denoised predictions (second column) and high SNR patch (last column).}
\label{fig:HDN_perf_CCPsvsMT}
\end{figure}
}

\newcommand\figHDNFactinVsCCPs{
\begin{figure}[!tbp]
  \centering
  \begin{minipage}[b]{0.32\textwidth}
    \includegraphics[width=\textwidth]{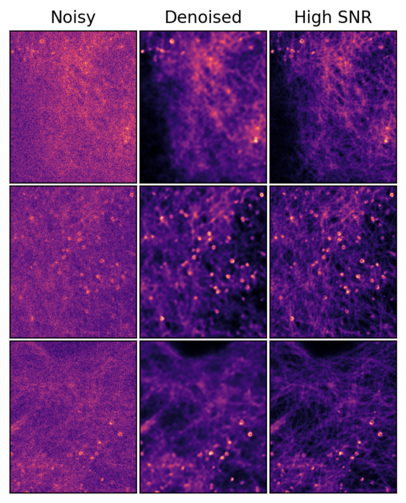}
  \end{minipage}
  \hfill
  \begin{minipage}[b]{0.32\textwidth}
    \includegraphics[width=\textwidth]{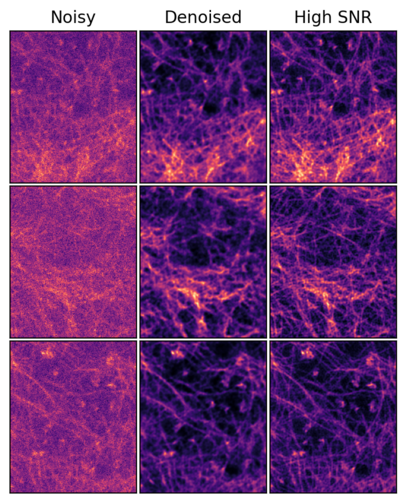}
  \end{minipage}
  \hfill
  \begin{minipage}[b]{0.32\textwidth}
    \includegraphics[width=\textwidth]{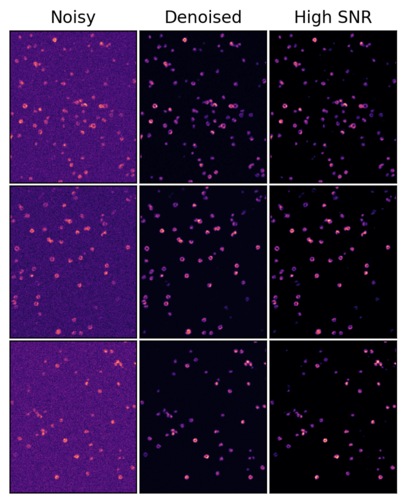}
  \end{minipage}
  
\caption{Qualitative performance of HDN on F-actin \vs CCPs task input and its two constituent channels. Left panel shows the denoising performance on input for our splitting task and central and right panel shows its denoising performance on its two constituent channels. Within each panel, we show three random patches (rows) of size $256\times213$. Specifically, we show the input (first column), denoised predictions (second column) and high SNR patch (last column).}
\label{fig:HDN_perf_FactinVsCCPs}
\end{figure}
}

\newcommand\figHDNFactinVsMT{
\begin{figure}[!tbp]
  \centering
  \begin{minipage}[b]{0.32\textwidth}
    \includegraphics[width=\textwidth]{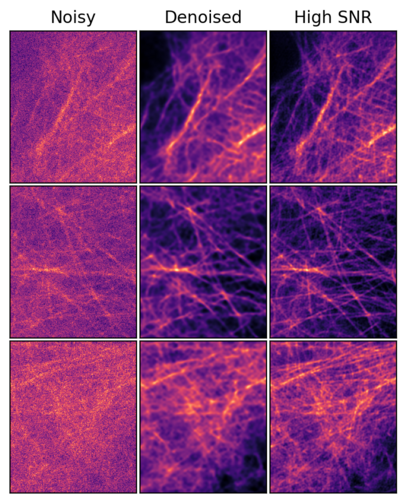}
  \end{minipage}
  \hfill
  \begin{minipage}[b]{0.32\textwidth}
    \includegraphics[width=\textwidth]{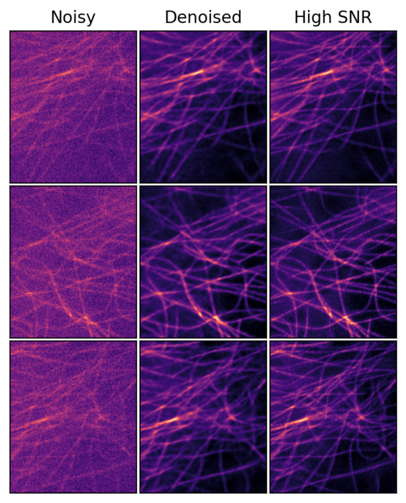}
  \end{minipage}
  \hfill
  \begin{minipage}[b]{0.32\textwidth}
    \includegraphics[width=\textwidth]{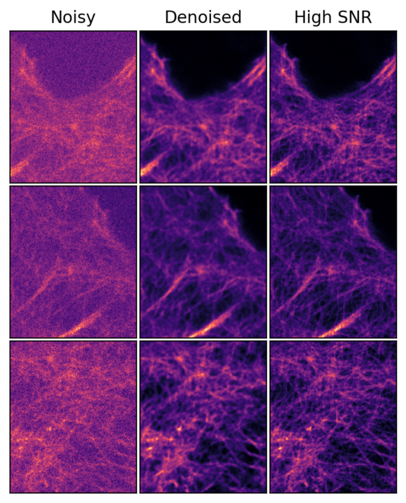}
  \end{minipage}
  
\caption{Qualitative performance of HDN on F-actin \vs MT task input and its two constituent channels. Left panel shows the denoising performance on input for our splitting task and central and right panel shows its denoising performance on its two constituent channels. Within each panel, we show three random patches (rows) of size $256\times213$. Specifically, we show the input (first column), denoised predictions (second column) and high SNR patch (last column).}
\label{fig:HDN_perf_FactinVsMT}
\end{figure}
}

\newcommand\figHDNFactinVsER{
\begin{figure}[!tbp]
  \centering
  \begin{minipage}[b]{0.32\textwidth}
    \includegraphics[width=\textwidth]{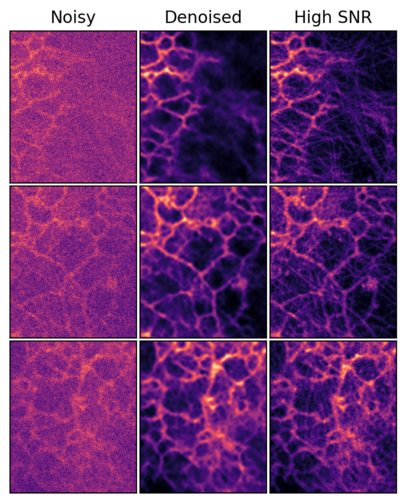}
  \end{minipage}
  \hfill
  \begin{minipage}[b]{0.32\textwidth}
    \includegraphics[width=\textwidth]{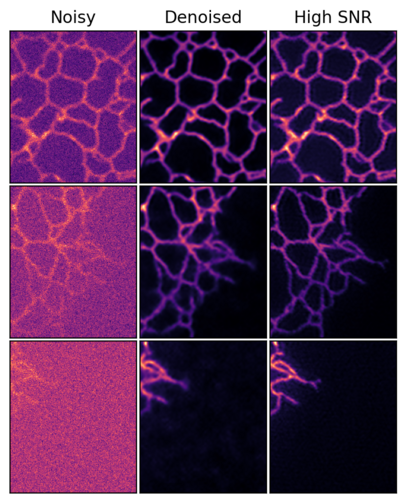}
  \end{minipage}
  \hfill
  \begin{minipage}[b]{0.32\textwidth}
    \includegraphics[width=\textwidth]{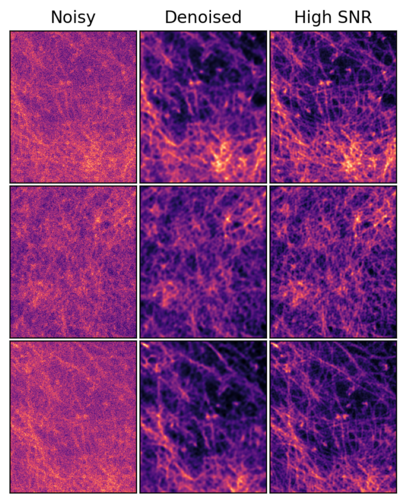}
  \end{minipage}
  
\caption{Qualitative performance of HDN on F-actin \vs ER task input and its two constituent channels. Left panel shows the denoising performance on input for our splitting task and central and right panel shows its denoising performance on its two constituent channels. Within each panel, we show three random patches (rows) of size $256\times213$. Specifically, we show the input (first column), denoised predictions (second column) and high SNR patch (last column).}
\label{fig:HDN_perf_FactinVsER}
\end{figure}
}

\newcommand\figPaviaATN{
\begin{figure}[!tbp]
  \centering
  \begin{minipage}[b]{0.49\textwidth}
    \includegraphics[width=\textwidth]{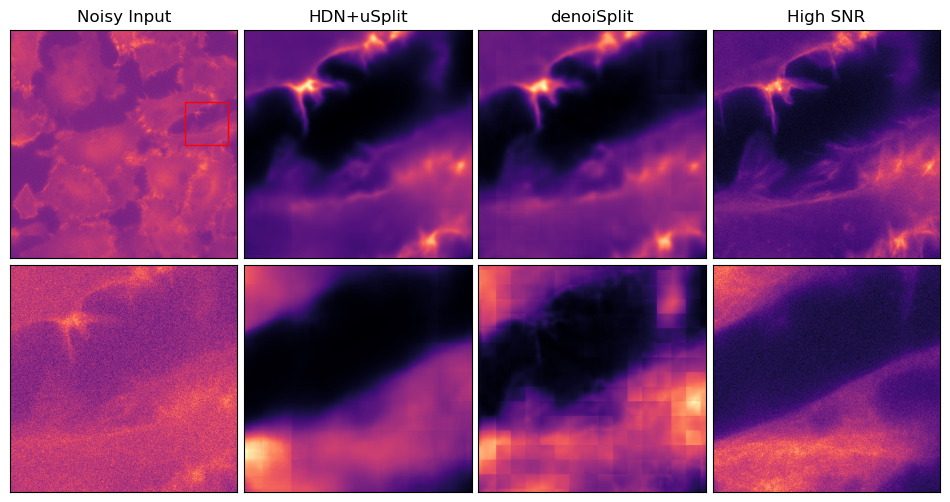}
  \end{minipage}
  \hfill
  \begin{minipage}[b]{0.49\textwidth}
    \includegraphics[width=\textwidth]{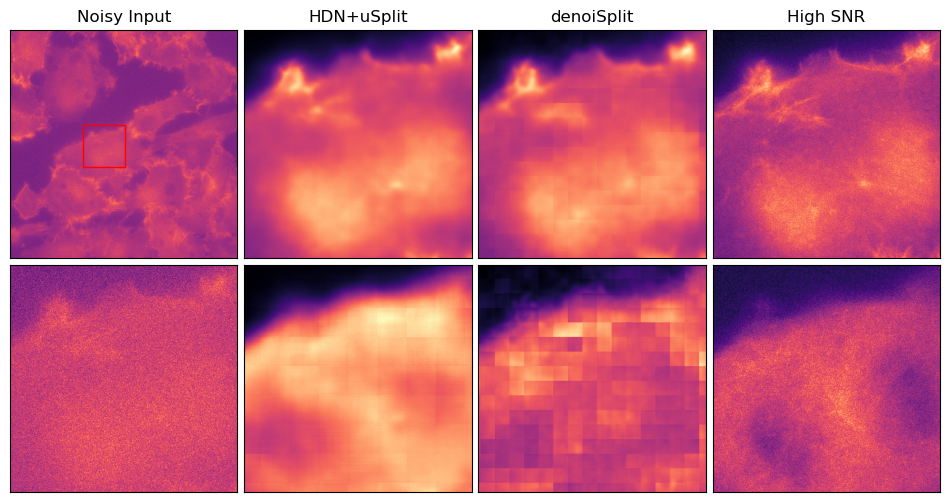}
  \end{minipage}

  \begin{minipage}[b]{0.49\textwidth}
    \includegraphics[width=\textwidth]{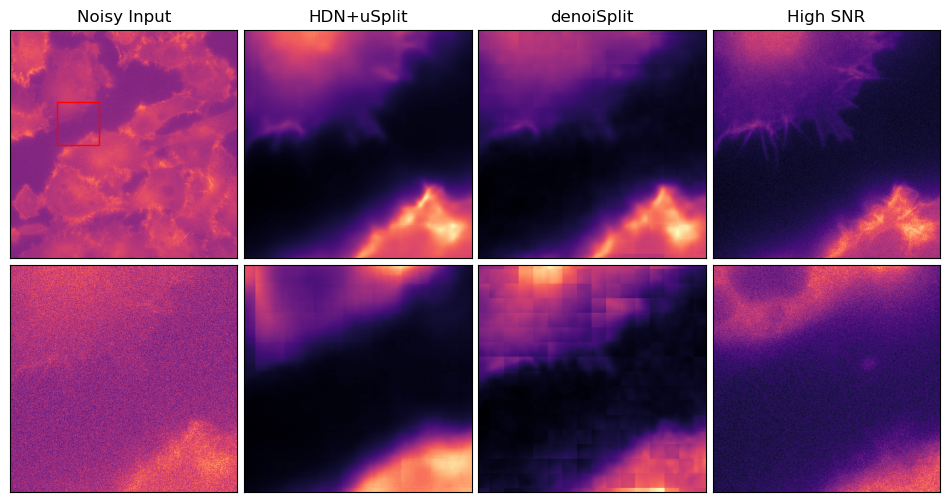}
  \end{minipage}
  \hfill
  \begin{minipage}[b]{0.49\textwidth}
    \includegraphics[width=\textwidth]{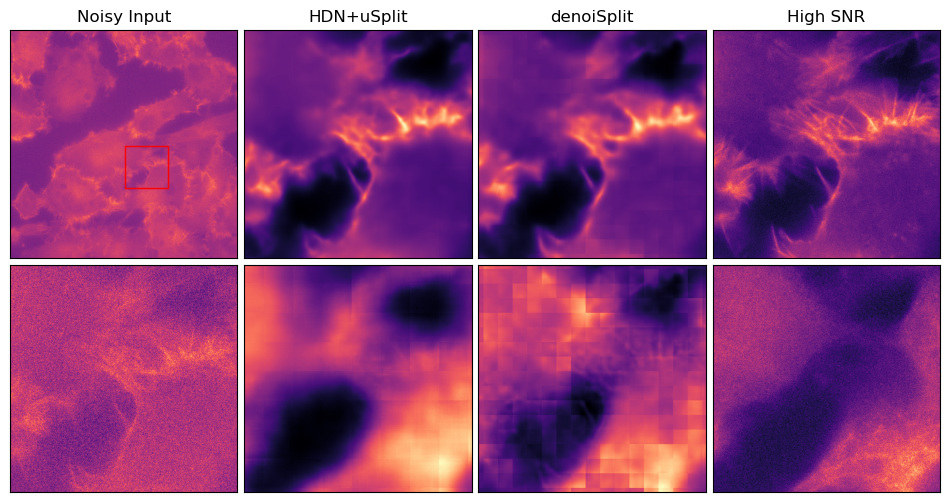}
  \end{minipage}

  \begin{minipage}[b]{0.49\textwidth}
    \includegraphics[width=\textwidth]{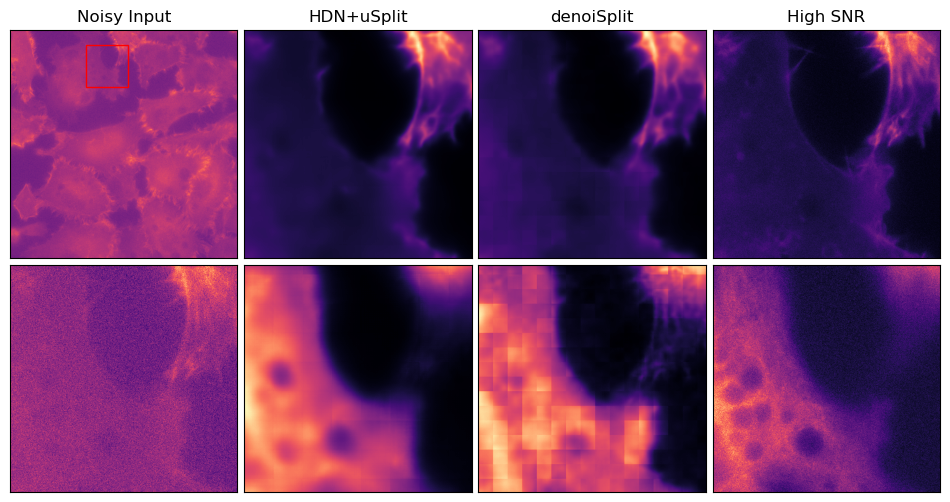}
  \end{minipage}
  \hfill
  \begin{minipage}[b]{0.49\textwidth}
    \includegraphics[width=\textwidth]{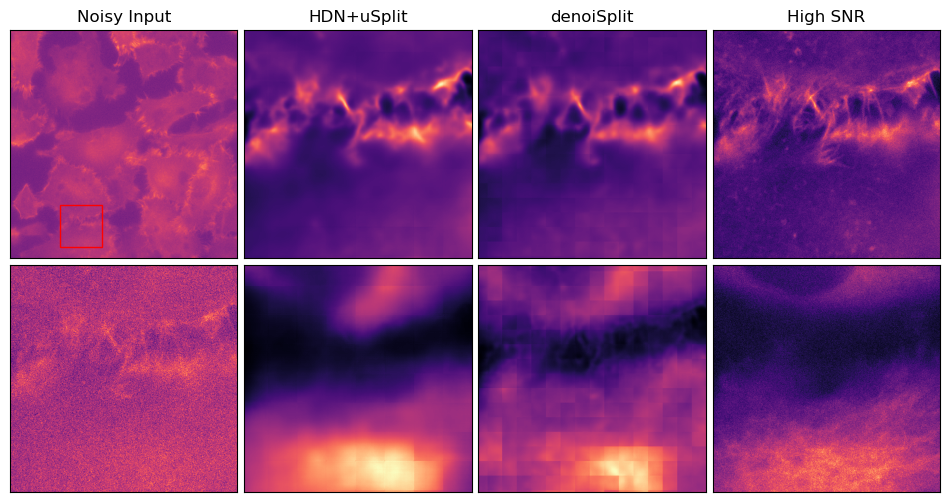}
  \end{minipage}

\caption{\textbf{PaviaATN Actin \vs Tubulin task} Here, we show performance of \denoiSplit and \hdnusplit for six random input patches of size $500\times500$ in six panels. Within each panel, we show the full input frame and its crop for which we do the predictions (column one). Next two columns have the predictions of \hdnusplit and \denoiSplit respectively. The last column is the high SNR ground truth. We observe that the splitting performance of both \hdnusplit and \denoiSplit does not reach the quality at which microscopists would find it useful. Between \hdnusplit and \denoiSplit, we see more tiling artefacts for \denoiSplit.}
\label{fig:paviaATN_bad_perf}
\end{figure}
}

\newcommand\figDeraining{
\begin{figure}[h]
\centering
\includegraphics[width=0.96\textwidth]{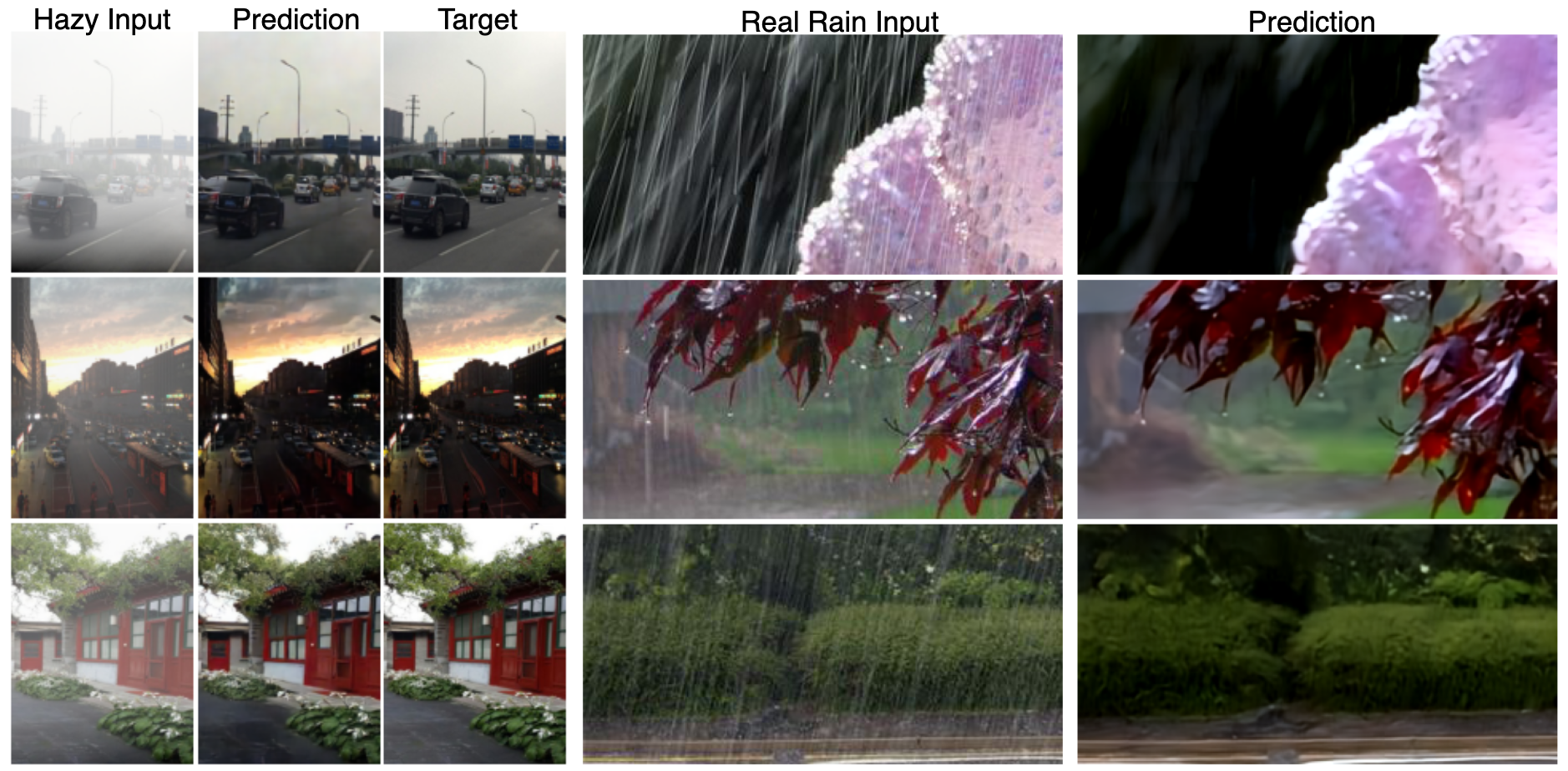} 
\caption{Qualitative results of \denoiSplit on De-hazing (Haze4K dataset) and De-raining (Trained on Rain100H dataset) tasks.}
\label{fig:derainng_dehazing}
\end{figure}
}

\begin{document}

\title{denoiSplit: a method for joint microscopy \\image splitting and unsupervised denoising} 

\titlerunning{denoiSplit}

\author{Ashesh\inst{1}\orcidlink{0000-0003-3778-0576} \and
Florian Jug\inst{1}\orcidlink{0000-0002-8499-5812}}

\authorrunning{Ashesh, F.\ Jug}

\institute{
Fondazione Human Technopole, Viale Rita Levi-Montalcini~1, 20157 Milan, Italy
\email{ashesh276@gmail.com; florian.jug@fht.org}}

\maketitle

\begin{abstract} 
In this work, we present \denoiSplit, a method to tackle a new analysis task, \ie the challenge of joint semantic image splitting and unsupervised denoising. 
This dual approach has important applications in fluorescence microscopy, where semantic image splitting has important applications but noise does generally hinder the downstream analysis of image content.
Image splitting involves dissecting an image into its distinguishable semantic structures. 
We show that the current state-of-the-art method for this task struggles in the presence of image noise, inadvertently also distributing the noise across the predicted outputs.
The method we present here can deal with image noise by integrating an unsupervised denoising subtask. 
This integration results in improved semantic image unmixing, even in the presence of notable and realistic levels of imaging noise.
A key innovation in \denoiSplit is the use of specifically formulated noise models and the suitable adjustment of KL-divergence loss for the high-dimensional hierarchical latent space we are training. 
We showcase the performance of \denoiSplit across multiple tasks on real-world microscopy images.
Additionally, we perform qualitative and quantitative evaluations and compare the results to existing benchmarks, demonstrating the effectiveness of using \denoiSplit: a single Variational Splitting Encoder-Decoder (VSE) Network using two suitable noise models to jointly perform semantic splitting and denoising. 
\end{abstract}

\section{Introduction}
\label{sec:intro}
Fluorescence microscopy remains a cornerstone in the exploration of cellular and sub-cellular structures, enabling scientists to visualize biological processes at a remarkable level of detail~\cite{Ghiran2011-np,Shroff2024-lm}. 
However, the ability to distinguish and analyze multiple structures within a single sample requires a multiplexed imaging protocol that requires extra time and effort~\cite{Shroff2024-lm}.
To address these downsides and enable for more efficient and new types of investigation, a powerful method for semantic image splitting was recently introduced~\cite{Ashesh2023-wtf}.

\figTeaser

Building on this previous work~\cite{Ashesh2023-wtf}, we address a key challenge that persisted: noise in microscopy images and its adverse effect on the quality of image-splitting predictions. 
Recognizing the need for a method capable of handling noisy input images while maintaining the integrity of the semantic splitting task, we introduce a technique that not only builds on the strengths of \muSplit~\cite{Ashesh2023-wtf} but also incorporates unsupervised denoising capabilities, for example as in~\cite{Krull2020-my,Prakash2019-jw,Prakash2020-wr,Prakash2021-dz}. 
Figure~\ref{fig:teaser} outlines the overall approach we are proposing.

Together, these ingredients lead to a new method \denoiSplit. 
It refines the process of image decomposition, ensuring that even under high levels of pixel noises present in the entire body of available training data, the semantic integrity of the semantically split image components (the predictions) is well preserved. 
Additionally, \denoiSplit can assess data uncertainty by sampling from the learned posterior of possible splitting solutions, followed by evaluating the inter-sample variability. 
In Section~\ref{sec:our_approach}, we show how to use this possibility to predict the expected error \denoiSplit makes on a given input.

In summary, we believe that this work will open new avenues for the efficient and detailed analysis of complex biological samples, for example, in the context of fluorescent microscopy.
\section{Related Work}
\label{sec:related_work}
\subsection{Image Denoising}
Image denoising is a task that has a long and exciting history. 
Classical methods, such as Non-Local Means~\cite{Buades2005-jk} or BM3D~\cite{Dabov2007-bm3d}, were frequently and very successfully used before neural network based approaches have been introduced towards the end of the last decade~\cite{Weigert2017-hc,Zhang2017-cz,Weigert2018-pi,Krull2018-cj}. 

The advent of deep learning saw people exploit different aspects of noise and the way networks learn to enable denoising. 
Noise, while usually undesired, is simultaneously much harder to predict, as was elegantly demonstrated in~\cite{Ulyanov2020-tl}, leading to a zero-shot denoiser. 
In the case specific to pixel noises, \ie all forms of noise that are independent per image pixel (given the signal at that pixel)~\cite{Shroff2024-lm}, the impossibility to predict the noise was exploited in various ways, leading to important contributions such as Noise2Void~\cite{Krull2018-cj}, Noise2Self~\cite{Batson2019-ae}, or Self2Self~\cite{Ko2023-dg}. 
Another well known approach close to this family of approaches is Noise2Noise~\cite{Lehtinen2018-ba}, capable of denoising even more complex noises that can correlate beyond the confines of single pixels.

\sloppy

To further improve denoising performance, Probabilistic Noise2Void~\cite{Krull2020-my,Prakash2019-jw} introduced, and DivNoising~\cite{Prakash2020-wr} and Hierarchical DivNoising (\HDN)~\cite{Prakash2021-dz} reused the idea of suitably measured or trained pixel noise models.
Such noise models are, in essence, a collection of probability distributions mapping from a true pixel intensity to observed noisy pixel measurements (and vice versa). 

\fussy 

\subsection{Image Decomposition} 
Image decomposition is the inverse problem of splitting a given input image that is the superposition (\ie the pixel-wise sum) of two constituent image channels.
While the sum of two values is not uniquely invertible, if for each summand a prior on its value exists, even a unique solution can exist.
In a similar vein, having learned structural priors of the appearance of the two constituent image channels, an input image (a grid of observed pixels that are each a sum of two values) can be split into two pixel grids such that each one satisfies the respective structural prior.
In computer vision, reflection removal, dehazing, deraining \etc are some of the applications~\cite{Gandelsman2019-nn,Dekel2017-os,Bahat2016-qp,Berman2016-nz} for which image splitting can be used. 

More recently, image splitting in fluorescence microscopy was receiving heightened attention, probably because of the direct applicability and potential utility that a well-working approach can bring to this microscopy modality, which finds wide-spread use in biological investigations. 
In particular, a method called \muSplit~\cite{Ashesh2023-wtf} demonstrated impressive image splitting performance on several datasets, suggesting that it is ready to be used in biological research projects.

However, \muSplit requires relatively noise-free data for training and prediction, which limits its potential utility (see also Section~\ref{sec:experiments} or Figure~\ref{fig:fullframepredictions}).

\subsection{Uncertainty Calibration} 
The ability to correctly assess the quality of predictions is naturally useful.
Ideally, a predictive system capable of co-predicting a confidence value has the property that the predicted confidence scales with the average error of the prediction.
If the relationship between error and confidence is close to the identity, we call the uncertainty predictions of this system \textit{calibrated}. 

Early works tried to use the deviation of the prediction as a proxy of the network's confidence in the prediction~\cite{Weigert2018-pi}. 
Other works tried to calibrate the predicted standard deviation with the expected error (\ie the RMSE)~\cite{calibration_levi}. 
Earlier calibration works were mainly concerned with classification tasks~\cite{calib_platt}. 
However, in~\cite{calibration_levi}, these approaches were reformulated in the context of regression.

In~\cite{calibration_levi}, the authors propose a way to evaluate calibration. 
They train a separate branch to predict a standard deviation per pixel that expresses its prediction uncertainty. 
For evaluating the calibration quality, the authors clustered examples on the basis of the predicted standard deviation values. 
Within each cluster, the predicted uncertainty is then compared with the empirical uncertainty (RMSE loss). 
To further improve the calibration, the authors propose a simple, yet effective scaling methodology wherein they learn a scalar parameter on re-calibration data, \ie a subset of data not included in the training data. 
(In our experiments, we use the validation data for this purpose.)
This scalar gets multiplied to the predicted uncertainty values, which then reduces the calibration error.
\section{Problem Formulation}
\label{sec:problem_formulation}
Lets denote a noise free dataset containing $n$ pairs of images as $D=(C_1,C_2)$, with each $C_i$ containing $n$ images $C_i=(c_{i,j} | 1\leq j\leq n)$.
Lets define a corresponding set of images $X=(x_j | 1\leq j\leq n)$, such that all $x_j=c_{1,j}+c_{2,j}$ are the pixel-wise sum of the two corresponding channel images. 

Although $D$ is typically not available (or even observable), in practice we can only observe noisy data, denoted here by $D^N=(C_1^N,C_2^N)$. 
Analogously to before, we define $X^N=(x_j^N | 1\leq j\leq n)$, such that $x_j^N=c_{1,j}^N+c_{2,j}^N$ are the pixel-wise sum of the noisy channel observations.

Given one $x_j^N\in X^N$ of $D^N$, the task at hand is to predict the noise free and unmixed tuple $(c_{1,j},c_{2,j})$.
We shall denote the predictions made by a trained \denoiSplit network by $(\hat{c}_{1,j},\hat{c}_{2,j})$.

Whenever above notions are used in a context that makes the $j$ in the subscript redundant, we allow ourselves to omit them for brevity and readability.

For evaluation purposes, we will in later sections use high-quality microscopy datasets that contain minimal levels of noise as surrogates for $D$, $X$, $C_1$, and $C_2$, but we never use them during training, and only their noisy counterparts are used. 
\section{Our Approach}
\label{sec:our_approach}
In the following sections we describe the main ingredients of \denoiSplit, namely 
the hierarchical network structure we use (Section~\ref{sec:approach_architecture}), the changed loss term for variational training of the splitting task (Section~\ref{sec:approach_kl_loss}),
the noise models we employ to enable the joint unsupervised denoising (Section~\ref{sec:approach_noise_models}),
and an uncertainty calibration methodology allowing us to estimate the prediction error introduced by aleatoric uncertainty in a given input image (Section~\ref{sec:approach_calibration}).
\subsection{Network Architecture and Training Objective}
\label{sec:approach_architecture}
In this work, we employ an altered Hierarchical VAE (\HVAE) network architecture.
\HVAEs were originally described in~\cite{Sonderby2016-lq} and later adapted for image denoising in~\cite{Prakash2021-dz} and for image splitting in~\cite{Ashesh2023-wtf}.
In general terms, \HVAEs learn a hierarchical latent space, with the lowest hierarchy level encoding detailed pixel-level structure, while higher hierarchy levels capture increasingly larger scale structures in the training data.

For \denoiSplit, we modify the \HVAE architecture so that it no longer remains an autoencoder.
Instead, our outputs are the two unmixed channel images $(\hat{c}_1,\hat{c}_2)$, motivating us to call the resulting architecture a Variational Splitting Encoder-Decoder (VSE) Network (see~\cref{fig:teaser}).

Our objective is to maximize the likelihood over the noisy two channel dataset we train on, \ie, finding decoder parameters $\decparams$ such that 
\begin{equation}
    \decparams = \argmax_{\decparams} \sum_{1\leq j\leq n} \log P(c_{1,j}^N, c_{2,j}^N;\decparams).
\end{equation}

Using the modified evidence lower bound (ELBO), as proposed in~\cite{Ashesh2023-wtf} and assuming conditional independence of the two predictions $(\hat{c}_{1},\hat{c}_{2})$ given the latent space embedding, we maximize
\begin{equation}
        \:E_{q(z|x;\encparams)} [
        \log P(c^N_1|z;\decparams) + \log P(c^N_2|z;\decparams)
        ] \\
        -KL(q(z|x;\encparams),P(z)),
    \label{eq:usplit_loss}
\end{equation}

where $q(z|x;\encparams)$ is the distribution parameterized by the output of the encoder network $\text{Enc}_{\encparams}(x)$, $P(c^N_i|z;\encparams)$ is the distribution parameterized by the output of the decoder network $\text{Dec}_{\decparams}(z)$ and $KL()$ denotes the Kullback-Leibler divergence loss. 
As in~\cite{Ashesh2023-wtf}, $P(z)$ factorizes over the different hierarchy levels in the network. 
Details about training, hyperparameters, \muSplit, and its relationship with \HDN and \denoiSplit can be found in \supsection{2}.

Similar to the way noise models had been employed in the context of denoising~\cite{Prakash2021-dz}, we model the two log likelihood terms $\log P(c^N_1|z;\decparams)$ and $\log P(c^N_2|z;\decparams)$ using noise models which we describe in detail in Section~\ref{sec:approach_noise_models} and our open code repository\footnote{https://github.com/juglab/denoiSplit}.
\subsection{Hierarchical KL Loss Weighing for Variational Training}
\label{sec:approach_kl_loss}
In \muSplit, the authors showed SOTA performance on a multitude of splitting tasks. 
However, used datasets were close to noise-free, making the task at hand simpler then the one we outlined in Section~\ref{sec:problem_formulation}. 
When \muSplit is trained on noisy datasets, the resulting channel predictions are themselves noisy. 
After analyzing this matter, we concluded that a modified KL loss can help reduce the amount of noise reconstructed by the decoder.

In more technical terms, let $Z$ be a hierarchical latent space and $Z[i]$ denote the latent space embedding at $i$-th hierarchy level, having shape $(c, h_i, w_i)$, with $c$ being channel dimension, and $h_i$, $w_i$ the height and width of the latent space embedding.
Now let $\text{KL}_i$ denote the KL-divergence loss tensor computed on $Z[i]$, which has the same shape as $Z[i]$ itself. 

In \muSplit, the corresponding scalar loss term $\text{kl}_i$ is defined as \(\text{kl}_i = \alpha\cdot \sum_{j,h,w}\frac{\text{KL}_i[j,h,w]}{ h_i\cdot w_i}\),
with $\alpha$ being a suitable constant. 
Observe that the denominator makes each $\text{kl}_i$ be the average of all values in $\text{KL}_i$, making the respective values not scale with the size of $Z[i]$, even though lower hierarchy levels ($Z[i]$ for smaller $i$) have more entries.
However, this also means that the KL loss for the individual pixels in these lower hierarchy levels is given less weight. 
Hence, smaller structures, such as noise itself, can more easily seep through such pixel-near hierarchy levels. 

In this work, we diverge from this formulation and return to a more classical setup where we compute the scalar loss term for the $i$-th hierarchy level $Z[i]$ as

\begin{equation}
    \text{kl}_i = \alpha\cdot \sum_{j,h,w}\text{KL}_i[j,h,w]. 
\label{eq:denoisplitkl}
\end{equation}

The decisive difference is that this changed formulation gives more weight to the KL loss at lower hierarchy levels, leading to more strongly enforcing the Gaussian nature the KL loss enforces, and therefore hindering noise from being as easily represented during training. 
We refer to this architecture as \usplitKL and show qualitative and quantitative results in Section~\ref{sec:experiments} and Tables~\ref{tab:overallperf}.

The next section extends on \usplitKL by adding unsupervised denoising, adding the last ingredient to the \denoiSplit approach we present in this work.
\subsection{Adding Suitable Pixel Noise Models}
\label{sec:approach_noise_models}
As briefly introduced in Section~\ref{sec:related_work}, pixel noise models are a collection of probability distributions mapping from a true pixel intensity to observed noisy pixel measurements (and vice-versa)~\cite{Krull2020-my}.
They have previously been successfully used in the context of unsupervised denoising~\cite{Prakash2020-wr,Prakash2021-dz} and we intend to employ them for this purpose also in the setup we are presenting here.
We use the fact that, given a measured (noisy) pixel intensity, a pixel noise model returns a distribution over clean signal intensities and their respective probability of being the underlying true pixel value.

We incorporate this likelihood function into the loss of our overall setup, encouraging \denoiSplit to predict pixel intensities that maximize this likelihood and thereby values that are consistent with the noise properties of the given training data.

Since \denoiSplit, in contrast to existing denoising applications, predicts two images (the two unmixed channels), we employ two noise models and add two likelihood terms to our overall loss.

More formally, in VAEs~\cite{Kingma2019-mx} and \HVAEs, the generative distribution over pixel intensities is modeled as a Gaussian distribution with its variance either clamped to $1$ or also learned and predicted.
We change our VSE Network to only predict the true pixel intensity and replace the Gaussian distribution mentioned above by the distributions defined in two noise models $\mathit{P}^\text{nm}_1(c^N_1|c_1)$ and $\mathit{P}^\text{nm}_2(c^N_2|c_2)$, one for each respective unmixed output channel. 
These noise models are pixel-wise independent, \ie, 
\begin{equation}
    \mathit{P_i}^\text{nm}(c^N_i|c_i) = \prod_k \mathit{P}^\text{nm}_i(c^N_{i}[k]|c_{i}[k]),\text{  } i\in\{1,2\},
\end{equation}
where $c^N_{i}[k]$ is the noisy pixel intensity for the $k$-th pixel and $c_{i}[k]$ the corresponding noise-free intensity value. 
This independence makes them particularly suitable for microscopy data where Poisson and Gaussian noise are the predominant pixel noises one desires to remove.

Since we now directly predict the noise-free pixel values, the output of the decoder can directly be interpreted as 
$\text{Dec}_{\decparams}(z) = (\hat{c}_1,\hat{c}_2)$ and the total loss for \denoiSplit now becomes
\begin{equation}
        \:E_{q(z|x;\encparams)} [
        \log P^{\text{nm}}(c^N_1|\hat{c}_1) + \log P^{\text{nm}}(c^N_2|\hat{c}_2)
        ] \\
        -KL(q(z|x;\encparams),P(z)).
    \label{eq:our_loss}
\end{equation}

In~\cite{Prakash2020-wr}, two ways for the creation of noise models are described, and the decision to pick which method depends upon whether or not one has access to the microscope from which data was acquired. 
In \supsection{5}, we describe the process of noise model generation and also compare performance between these two methodologies.
\subsection{Computing Calibrated Data Uncertainties}
\label{sec:approach_calibration}
The idea of calibration is for those network setups that produce both prediction and a measure of uncertainty for the prediction. 

Networks that can co-assess the uncertainty of their predictions are called calibrated, when the predicted uncertainties are in line with the measured prediction error.
To improve the calibration of a given system, one can find a suitable transformation from uncertainty predictions to measured errors (\eg, the RMSE).
After such a transformation is found, an ideal calibrated plot would be tightly fitting $y=x$, with y and x being the error and estimated uncertainty, respectively. See, for example, Figure~\ref{fig:sampling}. 
Since VSE networks, similar to \VAEs, are variational inference systems, we can sample from their latent encoding and thereby sample from an approximate posterior distribution of possible solutions giving us the data uncertainty.
In this section, our intention is to utilize this ability to predict a reliable uncertainty term for our results.

For this, we adapt the calibration methodology of~\cite{calibration_levi}. 
In contrast to the approach described there, we propose to use the variability in posterior samples to estimate a pixel-wise standard deviation. 
More specifically, we sample $k=50$ predictions for each input image and compute the pixel-wise standard deviations $\sigma_1$ and $\sigma_2$ for the two predicted image channels $\hat{c}_1$ and $\hat{c}_2$, respectively. 
This gives us uncertainty predictions. 

Next, we calibrate these uncertainty predictions by scaling them appropriately with the help of two learnable scalars, $\alpha_1$ and $\alpha_2$.  
Following~\cite{calibration_levi}, we assume that pixel intensities come from a Gaussian distribution. 
The mean and standard deviation of this distribution are the pixel intensities of the MMSE prediction, \ie the image obtained after averaging $k=50$ predictions, and the scaled $\sigma$, respectively. 
We learn the scalars $\alpha_1$ and $\alpha_2$ by minimizing the negative log-likelihood over the recalibration dataset. 
It is important to note that the presented calibration procedure does not alter the original predictions but instead learns a mapping that best predicts the measured error.

To evaluate the quality of the resulting calibration, we sort the scaled standard deviations $\sigma_i\cdot s_i$ for each pixel in a predicted channel and build a histogram over $l=30$ equally sized bins $B_i^j$.
We then compute the root mean variance (RMV) and RMSE for each bin $j$ and channel $i$ as\newline
\begin{minipage}{.5\textwidth}
\begin{equation*}
    \text{RMV}_i(j) = \sqrt{\frac{1}{|B^j_i|}\sum_{k\in B^j_i} (\sigma^k_i\cdot \alpha_i)^2}
\end{equation*}
\end{minipage}%
\begin{minipage}{.5\textwidth}
\begin{equation*}
    \text{RMSE}_i(j) = \sqrt{\frac{1}{|B^j_i|}\sum_{k\in B^j_i} (c_i[k] - \hat{c}_i[k])^2}
\end{equation*}
\end{minipage}
As in Section~\ref{sec:problem_formulation}, $c_i[t]$ and $\hat{c}_i[t]$ denote the noise-free pixel intensity and the corresponding prediction for $i$-th channel. 
In \cref{fig:sampling}, we plot the RMSE vs.\ RMV for multiple tasks, observing that the plots closely resemble the identity $y=x$.  
Following~\cite{calibration_levi}, we use the validation dataset for recalibration and show calibration plots on the test dataset. 
\figBioSR
\figSampling
\figPatchComparison
\section{Experiments and Results}
\label{sec:experiments}
\subsection{Datasets}
\label{sec:datasets}
\paragraph{BioSR dataset} We work primarily with BioSR dataset~\cite{Jin2023.02.03.526797}, a comprehensive dataset comprising fluorescence microscopy images of multiple cell structures.
For our experiments, we have picked four structures, namely clathrin-coated pits (CCPs), microtubules (MTs), endoplasmic reticulum (ER), and F-actin. 

Since the raw data quality is very high and only a small amount of image noise is present in the individual micrographs, we add Gaussian noise and Poisson noise of various levels to these raw data.
The artificially noisy images are used to train \denoiSplit, while the raw data is shown to convince the reader of the validity of our approach and to compute evaluation metrics (see~\cref{fig:fullframepredictions,fig:patchcomparison} and~\cref{tab:overallperf}). 

\paragraph{Hagen et al. Actin-Mitochondria Dataset} 
We picked the noisy Actin and Mitochondria channels from Hagen et al.~\cite{Hagen2021-xh}, channels having real microscopy noise. 
For evaluation, we use the corresponding high-SNR (noise-free) channels provided in the dataset.

\paragraph{Synthetic Noise Levels}
We work with $4$ levels of zero-mean Gaussian noise and two levels of Poisson noise. 
For Gaussian noise, we compute the standard deviation of the input data $X^N$ for each of the tasks and scale the noise relative to one standard deviation. 
Specifically, the $4$ scaling factors are $\{1, 1.5, 2, 4\}$. 
In cases where Poisson noise is added, and since it is already signal dependent, we use a constant factor of $1000$ to hit a realistic-looking level of Poisson noise. We also consider the case where Poisson noise is not added, which we denote by the Poisson level of $0$ in~\cref{tab:overallperf}.
To remove any remaining room for misinterpretations, we provide a pseudo-code for the synthetic noising procedure in the \supsection{2}.

\subsection{Baselines}
\label{subsec:baselines}
We conducted all experiments with two baseline setups, \muSplit and \hdnusplit.
In the original \muSplit work~\cite{Ashesh2023-wtf}, the authors introduce three architectures, each with a different trade-off between GPU efficiency, speed and performance. 
We pick the most balanced variant, \HVAE + \regularLC, which we refer to as \muSplit.

The second baseline, to which we refer to as \hdnusplit, is a sequential application of Hierarchical DivNoising (\HDN), one of the leading unsupervised denoising methods for microscopy datasets~\cite{Prakash2021-dz}, and the \muSplit setup from above. 
We first denoise all input images $x_i^N$ and the respective two channel images $c_{1,j}^N,c_{2,j}^N$.
Note that each set of the three kinds of image are denoised with a separately and specifically trained \HDN. 

Next, we use the denoised predictions of $X^N, C_1^N, C_2^N$ to train a \muSplit network as we did for the first baseline. The expectation from this baseline is to give denoised splitting results, which we also show in~\cref{fig:patchcomparison}. 
We show the denoised \HDN predictions in the supplement. 
\tabOverallPerf
\realNoisePerformance
\subsection{Qualitative and Quantitative Evaluation of Results}
\label{sec:qualitative_evaluation}
We show the quality of results our methods can obtain in~\cref{fig:fullframepredictions,fig:sampling,fig:patchcomparison,fig:realnoise}. 

In~\cref{fig:fullframepredictions}, we show predictions on full input images ($960\times 960$ pixels).
We can see that \muSplit does unmix the given inputs and even partially reduces the noise.
Still, the results by \denoiSplit have a much higher resemblance to the high-SNR microscopy images shown in the rightmost column, even though they have never been presented during training (which was conducted only on noisy images, as shown in the second column).
In~\cref{fig:realnoise}, we show the results on Hagen et al.~\cite{Hagen2021-xh} dataset. Again, we observe \denoiSplit outperforming \muSplit and \hdnusplit. Please refer to~\supsection{2} for more details.   

In~\cref{fig:sampling}, we show zoomed $256\times 256$ portions of full predictions to allow the reader to also appreciate the prediction quality of smaller structures contained in the data.
Furthermore, we show two posterior samples ($S_1$ and $S_2$) and their highlighted differences ($S_1-S_2$).
The second to last column shows the average of $50$ posterior samples (the approximate MMSE~\cite{Prakash2020-wr}).
We show the calibration plot in the second row, first column of every panel where we can see a clear predictive (and close to linear) behavior of RMSE from RMV.

In~\cref{fig:patchcomparison}, we also show $256\times 256$ insets on inputs and results by \hdnusplit and \denoiSplit, showing that the fine details are better preserved by our proposed method.
Note that these very details make all the difference when such methods are used on fluorescence microscopy data for the sake of downstream analysis.

For quantitative quality evaluations, we use the well known and established PSNR and MS-SSIM metrics~\cite{Shroff2024-lm} to evaluate all the results of our experiments and report these results in~\cref{tab:overallperf}. 
These quantitative evaluations clearly show that our proposed methods improve considerably over \muSplit and the sequential denoising and splitting baseline \hdnusplit.
In \supsection{1 and 8}, we provide results on more splitting tasks, including one failure mode. In \supsection{3}, we show a proof-of-concept application to de-hazing and de-raining tasks. 
\section{Discussion and Conclusion}
\label{sec:discussion}
\label{sec:conclusion}
We present \denoiSplit, the first method that takes on the challenge of joint semantic image splitting and unsupervised denoising. 
This advancement in handling noise is crucial, considering the limitations microscopists face in acquiring high-SNR images, often due to practical constraints such as sample sensitivity and limitations in imaging technology. 
Unlike in a sequential approach of image denoising followed by training and applying \muSplit on denoised data, \denoiSplit streamlines the process into a single end-to-end model, which on the one hand reduces the complexity and computational resources required for training and inference, and on the other hand leads to better results.

One of our methodological contributions is integrating noise models, originally developed for unsupervised denoising task, into the image-splitting setup.
Given the fact that different microscope configurations produce images with different noise levels and Noise models can be made specific to each microscope configuration, the integration of noise models in our setup can go a long way in allowing a microscope-specific \denoiSplit setup thereby producing high-quality denoised and split predictions.  
Our work has additional interesting features, which we believe will improve its adoption among microscopists, \ie, it supports variational sampling and calibration, allowing microscopists to observe multiple solutions for a given input and also to have an estimate of error for every predicted pixel. 

In the future, we want to work on domain adaptation techniques to fine-tune existing models on noisy data from slightly different image domains or microscopy modalities. 
This will further ease the applicability of \denoiSplit for biomedical researchers.
This continuous development aims to bridge the gap between computational imaging methods and the practicalities and limitations of modern microscopy, benefiting our overall goal of elevating the rate of scientific discovery in the life sciences by conducting cutting-edge methods research.

\section*{Acknowledgements}
\label{sec:ack}
This work was supported by 
the European Commission through the Horizon Europe program (IMAGINE project, grant agreement 101094250-IMAGINE and AI4LIFE project, grant agreement 101057970-AI4LIFE) as well as the compute infrastructure of the BMBF-funded de.NBI Cloud within the German Network for Bioinformatics Infrastructure (de.NBI) (031A532B, 031A533A, 031A533B, 031A534A, 031A535A, 031A537A, 031A537B, 031A537C, 031A537D, 031A538A).
Additionally, the authors also want to thank Damian Dalle Nogare of the Image Analysis Facility at Human Technopole for useful guidance and discussions and the IT and HPC teams at HT for the compute infrastructure they make available to us.

\bibliographystyle{splncs04}
\bibliography{main}
\newpage

\begin{center}

  \textbf{\Large Supplementary Material denoiSplit:\\ a method for joint microscopy \\image splitting and unsupervised denoising}\\[.2cm]
Ashesh,
Florian~Jug \\
Jug Group, Fondazione Human Technopole, Milano, Italy, 
\\
\end{center}
\setcounter{equation}{0}
\setcounter{figure}{0}
\setcounter{table}{0}
\setcounter{page}{1}
\renewcommand{\thepage}{S.\arabic{page}} 
\renewcommand{\thefigure}{S.\arabic{figure}}
\renewcommand{\thetable}{S.\arabic{table}}
\renewcommand*{\thesection}{S.\arabic{section}}
\setcounter{section}{0}

\tabOverallPerfOtherDset
\section{Performance on more splitting tasks}
In this section, we train our models and baselines on four more tasks.  
We train two tasks from the BioSR dataset. 
Specifically, we add F-actin \vs CCPs and F-actin \vs Microtubules tasks. 
In~\cref{tab:overallperfOtherdset}, we present the quantitative evaluation on these two tasks. 
Similar to the results from Table 1 of the main manuscript, here as well, we find our methods, specifically \denoiSplit to outperform others in most cases. 

We worked on three additional joint denoising-splitting tasks from two other datasets which we describe next.
\paragraph{Hagen et al. Actin-Mitochondria Dataset} 
We picked the high-resolution Actin and Mitochondria channels from Hagen et al.~\cite{Hagen2021-xh} which were also used in~\cite{Ashesh2023-wtf}. 
Similar to our tasks from BioSR dataset, we added Gaussian and Poisson noise.

\paragraph{PaviaATN dataset}~\cite{Ashesh2023-wtf} We worked with the Actin and Tubulin channel provided by the dataset. It is worth noting that in terms of PSNR, we picked the hardest of the three tasks worked upon by Ashesh et al.~\cite{Ashesh2023-wtf}.
This is the task on which \denoiSplit and all the baselines perform poorly.
We discuss more on these results in~\cref{sec:limitations}. 

We provide the results on tasks generated from PaviaATN and Hagen et al. datasets in~\cref{tab:overallperfothertable}. 
We show the full-frame predictions for tasks not shown in the main manuscript in~\cref{fig:realnoise,fig:factinVsCCPs,fig:factinVsER,fig:factinVsMT}.

\section{Details on architecture, hyperparameters, training and evaluation}
\label{sec:hparam_details}
As stated in the main manuscript, our \denoiSplit and \usplitKL are built on top of \muSplit architecture. 
In addition to the major changes that are discussed in the main manuscript, we have enabled $\textit{free bits}$~\cite{freebitsvaekingma} parameter and have set $\textit{free bits}=1$. But similar to~\cite{Ashesh2023-wtf}, we also upper bounded the log of variance of the latent space to $20$ across all hierarchy levels for stability in training.


\denoiSplit, \usplitKL and \HDN have been trained with the learning rate of $0.001$, batch size of $32$, patch size of $128$, max epoch of $400$ and with $16$ bit precision. 
For every task, $80\%$ of the data was allocated as training data, $10\%$ of the data as validation data, and $10\%$ as the test data. 
For \muSplit baseline and for \muSplit used as part of \hdnusplit, we use the same training configuration as mentioned in~\cite{Ashesh2023-wtf}.  
For PSNR, we use the range invariant PSNR formulation which is commonly used in this field~\cite{Ashesh2023-wtf,Weigert2018-pi}. 
When working with Actin vs Mito task (Fig. 5 of the main text), we first scale the predictions in a way described in~\cite{Weigert2018-pi} and then use the Multiscale SSIM metric between the high-SNR groundtruth and the scaled prediction. 
The scaling is necessary because the high-SNR groundtruth has much higher pixel intensties than the low-SNR data on which the models are trained. 
Due to this the predictions also have lower pixel intensity values. 
We do not need to do this for PSNR separately because the version we use already does the scaling. 

We have provided code with this supplement where information about how Poisson noise is added is present in the file $\textit{vanilla\_dloader.py:L175}$. Specifically, we use numpy python package to add Poisson noise as 
\begin{verbatim}
data = np.random.poisson(data / poisson_factor) * poisson_factor    
\end{verbatim}
We use $\text{poisson\_factor}=1000$. 
Finally, in~\cref{tab:gaussianSigmaValues}, we provide the actual $\sigma$ values which were used to add Gaussian noise in different tasks.  

For the calibration plot shown in Figure 3 of the main text, 50 samples were used to estimate the RMV (Root mean variance).
\subsection{Architecture details of \muSplit}
For completeness, here we describe all relevant aspects of \muSplit~\cite{Ashesh2023-wtf}, the work that we built upon.
\muSplit was built by modifying a \HVAE framework and therefore inherited multiple hierarchy levels of latent spaces and the loss comprised of KL divergence and log-likelihood. 
In \muSplit, Ashesh et al.~\cite{Ashesh2023-wtf} removed the auto-encoding nature of the \HVAE framework by making the network predict two-channel output with input being a single channel.
They therefore had two components in their log-likelihood loss, one for each output channel.
In the log-likelihood loss of \muSplit, a pixel-wise variance was predicted along with the split prediction for each channel. 
Note that in \HVAE implementations, variance in the log-likelihood loss component is typically set to 1. 
\muSplit's prediction, on the other hand, is a 4-channel tensor, two channels being the prediction and the other two being the predicted pixel-wise log-variance.
As discussed in the main manuscript, one of our key contributions is that we integrated Noise models, originally developed for unsupervised denoising into the image decomposition task. 
We therefore did not need to predict the pixelwise log-variance.
Next, compared to the classical \HVAE formulation~\cite{Prakash2021-dz} of KL divergence loss, Ashesh et. al~\cite{Ashesh2023-wtf} relax the weights given to KL divergence loss components across different hierarchy levels, arguably to get better high-frequency details in the prediction. 
In our work, however, we find that this leads to a loss of denoising and sampling properties of the network, properties which \HVAEs typically have.
Therefore, as discussed in the main manuscript, we changed the KL loss weighting scheme used in \muSplit and adopted the formulation used in~\cite{Prakash2021-dz}. 
Ashesh et al.~\cite{Ashesh2023-wtf} also showed the theoretical soundness of their approach by deriving the ELBO loss for the image decomposition setup. 
Please refer to~\cite{Ashesh2023-wtf} for the proof.
Finally, one of their main contributions was the introduction of lateral contextualization (LC) wherein additional low-resolution images centered on the primary input patch, but covering larger spatial regions, were fed to the network through separate input branches. 
This enabled GPU-memory efficient assimilation of the information about the surrounding spatial context of the input patch. 
Since we primarily worked with the BioSR dataset which did not have as large structures as the ones used in~\cite{Ashesh2023-wtf}, we disabled the LC module and instead doubled the input patch size. 
\section{Applications on natural images}
While this work focuses on microscopy data, in this section, we briefly explore the utility of \muSplit to tasks on natural images. Specifically, we look at de-raining and de-hazing tasks. 
For de-hazing, we used Haze4K dataset~\cite{dehazing-dmtnet} and for de-raining, we used Rain100H dataset~\cite{derain-jorder}. We worked with the original version of Rain100H dataset containing 1800 clean/rainy training image pairs and 100 clean/rainy testing image pairs.
Due to the absence of pixel-independent noise in these datasets, we disabled the noise model in \denoiSplit. 

We present qualitative and quantitative results in Figure~\ref{fig:derainng_dehazing} and Tab.~\ref{tab:deraining},~\ref{tab:dehazing}. In the de-raining task, it was encouraging to observe that while training was done on images with synthetic rain, \denoiSplit was able to remove rain from real rainy images as well.
\figDeraining
\tableDeraining
\tableHazing

It should be noted that the absence of noise in these datasets did not allow the primary feature of our approach, unsupervised denoising using noise models, to be used in these tasks. 
However, we believe our approach holds more promise on tasks on natural images having a significant amount of noise.
\section{Practical relevance of \denoiSplit}
Here, we outline the intended usecase for our work. 
Similar to~\cite{Ashesh2023-wtf}, our work also aims to enable microscopists to extract multiple structures using a single fluorescence marker. 
There are two important practical considerations which our work addresses. 

Firstly, different microscopy projects, depending upon the nature of underlying specimen and microscope type, have different tolerances for the amount of laser power and the dwell time that can be used during acquisition.
This roughly translates to the amount of noise that will be present in the acquired micrographs. 
In this work, we cater to this necessity by working with different noise levels.  

Secondly, in most cases, there will be a need to purchase a single fluorescent marker that can bind to both cellular structure types one is interested in. 
Additionally, even after the purchase, it might still be challenging to get a decent staining of both structures with the marker. 
An imperfect staining can lead to under-expression of one of the two structure types in the imaged micrographs. 
Therefore, there is also an investment of time and expertise in getting a proper staining.
That being the case, it makes sense to first inspect the feasibility of the approach before making the investment of buying a new marker followed by getting the staining correct. 
Our method provides a way to get proof-of-concept splitting without making any of these investments.

We envision that microscopists should image individual channels in the same noise regime as is permitted in their project. 
They should then train the \denoiSplit and inspect the prediction quality. 
If there is room for adjustment in power and dwell time, they can re-acquire in a different noise regime and train the \denoiSplit again. 
Note that this acquisition can be done with a single-color setup and therefore can be done using their existing microscope configuration.

In case they find the model performance satisfactory in some feasible noise regime, they can then order the relevant fluorescent marker and can subsequently label their structures with it to get superimposed images. 
Finally, they need to finetune the trained network to this slightly different input data and then they can start using \denoiSplit. 
This problem of finetuning the model is outside the scope of this work and will be taken up in our future work. 

\denoiSplit also enables sampling which, as stated in the main manuscript, will enable microscopists to inspect the predictions to get a visual feeling about uncertain areas in prediction. To showcase this, we provide with this supplement, a \textit{sampling.gif} file where one could see 50 samples on a randomly choosen input from MT \vs ER task. 

\section{Noise model generation}
Depending upon the physical availability of the microscope,~\cite{Prakash2020-wr} described two ways in which the noise model can be generated. 
In this section, our motivation is to assess the performance difference that one should expect between these two ways. We find that different choices of noise model generation do not lead to very different performances. This result enables \denoiSplit to be used across a wide variety of scenarios. 
\subsection{Physical availability of Microscope}
The simpler and the better case is when we have access to the microscope which generated the noisy data. 
In this case, one needs to image the same content $N$ times which would result in acquisition of $N$ noisy versions of the same underlying specimen. 
The high SNR version is then computed by simply taking the pixel-wise average of these noisy samples. 
In this situation, for every clean signal value, one has $N$ noisy intensities. 
This data is then used to train the Gaussian-mixture based noise model. 

To simulate this condition, for every intensity value in the range $[0,65535]$, we obtain multiple noisy intensity values by adding the noise (Gaussian and/or Poisson) multiple times independently.  
\subsection{Physical unavailability of Microscope}
In case the microscope is not available, which is true for all publicly available microscopy datasets,~\cite{Prakash2020-wr} proposed a bootstrap noise model approach. 
In this method, the idea was to denoise the noisy data using some unsupervised/self-supervised denoising technique and use the noisy data and the predicted denoised data to generate the noise model.

To simulate this condition, we added noise to the noise free training data. 
This gave us the noisy data and clean data pair which we used to train the noise model. 
To showcase the applicability of our method to all publicly available datasets, we have generated all noise models using this approach.

It is worth noting that in the way described above, there is a source of error which has not been captured.
This is the error introduced by the denoising process for getting the clean data from its noisy counterpart. 
To assess the effect of denoising, we generate the noise model in yet another way. In this way, we added the noise to the noise free training data. 
We then use N2V~\cite{Krull2018-cj}, an off-the-shelf denoiser to denoise the images. We use the noisy data and the denoised images to generate the noise model.
\subsection{Performance comparison among different noise model generation procedures}
\figNoiseModelComparison
To assess how performance varies depending upon the methodology used for noise model creation which have been described above subsections, we train \denoiSplit three times, each time with its noise model computed using a different way.
In~\cref{fig:noiseModelMultipleWays}, we show the performance comparison. 
denoiSplit$+S_\infty$ simulates the case when one has access to a microscope. For every clean pixel, one has multiple corresponding noisy pixels.
denoiSplit$+S1$ denotes the case when one works with a noisy and the corresponding clean data pair. 
Note that, unlike the previous case, for every clean pixel, there is exactly one noisy pixel.

denoiSplit$+N2V$ denotes the case when one obtains the clean data after denoising with N2V. 
We train them on two tasks namely ER \vs. CCPs and ER \vs. MT over four Gaussian noise levels with Poisson noise ($\lambda=1000$) enabled. 
As can be observed from the plots, we don't see a large difference between the three approaches. 

This performance evaluation encourages \denoiSplit to be used on publicly available datasets and for proof-of-concept evaluations since it shows that having access to the microscope does not give large performance improvements and the \textbf{bottleneck in denoising-splitting is essentially the splitting task}. 
\figDenoisingSplittingComparison
\figCalibrationMmseVariation
\section{Quantifying model uncertainty}
In this section, we quantify how much the performance varies between multiple models trained independently on the same task under identical configuration.
For this, we picked ER vs MT task from BioSR dataset with $\sigma=1, \lambda=1000$. 
We trained the model 10 times and computed the PSNR and SSIM metrics on the test data. 
The mean and standard deviation of the PSNR values across different runs came out to be $29.84\pm0.18$ dB with individual PSNR values lying in $[29.7,30.3]$ dB.
For SSIM, it was $0.905\pm0.004$. 
Due to computing limitations, all the main text and the supplement tables use a single trained model per configuration to generate the metric values.  
\section{Comparison between denoising task and splitting task}
In this section, we compare between two computer vision tasks which are of relevance to us: (a) unsupervised denoising and (b) joint denoising-splitting. 
From~\cref{fig:denoisingVSsplitting}, we observe that unsupervised denoising is arguably a simpler task when compared to joint denoising-splitting. 
The PSNR between the prediction and high SNR micrographs is much better for \HDN as compared to \denoiSplit for most cases. 
We note that it is expected because in joint denoising-splitting, besides denoising, which is the sole task in unsupervised denoising, one needs to additionally do the job of image decomposition. 
But more interestingly, we observe that the prediction quality of one channel depends upon the other channel. 

We note that this observation is of considerable importance because it opens up the question of best pairing strategy: which two structures should be imaged by a single color fluorescent marker? 
This will be part of our future work.  


\section{On usefulness of using sampling for calibration}
In this section, we investigate our choice for estimating un-calibrated uncertainty using sampled predictions. 
For this, we estimate the uncalibrated pixelwise uncertainty using varying number of samples. 
We then follow our calibration procedure and learn the channelwise scalar to get the calibration plot. 
As can be observed in~\cref{fig:calibmmse}, as we increase the number of samples, the calibrated plot also improves thereby validating our choice.
\section{Qualitative evaluation of HDN denoising}
Here, we qualitatively evaluate the denoising behaviour of \HDN. 
We show three random input patches and the corresponding channel first and channel second crops from 6 different splitting tasks in~\cref{fig:HDN_perf_CCPsvsMT,fig:HDN_perf_ERvsCCPs,fig:HDN_perf_ERvsMT,fig:HDN_perf_FactinVsCCPs,fig:HDN_perf_FactinVsER,fig:HDN_perf_FactinVsMT}. 
We show the results on noisy dataset having Poisson($\lambda=1000)$ noise and Gaussian noise of relative scale $1.5$. 
Since main manuscript also shows the qualitative figures on this noise level, we believe this can be used together with the figures present in main manuscript to better understand the behaviour of \hdnusplit. 

\section{Failure cases: Avenues for future work}
\label{sec:limitations}
In this section, we inspect the worse performing cases in our work. 
One clear example is Actin \vs Tubulin task from PaviaATN~\cite{Ashesh2023-wtf} dataset whose quantitative evaluation is present in~\cref{tab:overallperfothertable}. However, as can be seen from~\cref{fig:paviaATN_bad_perf}, we find performance of both \hdnusplit and \denoiSplit to be unsatisfactory for it to be used by microscopists.   
We note a striking difference of this task with the tasks from BioSR dataset. 
Looking at $128\times128$ input patches for tasks from BioSR data, one could visually form an opinion about which structures in the input patch should belong to first channel and which to the second channel. 
In case of Actin \vs Tubulin task, we observe that making this opinion is much more difficult since local structures are much less discriminative.
It is only when looking at a larger context, one can form some opinion about which structures should be present in which channel. 
There seems to be another factor related to the nature of the structures which the channels are composed of. 
Informally speaking, individual channels have a "surface" like structure in Actin and Tubulin images of PaviaATN dataset as opposed to "curved lines", "mesh" and "dots" like structures in BioSR dataset. 

Between \denoiSplit and \hdnusplit, we observe that for this task, \denoiSplit has more tiling artefacts. It is not surprising for that to be the case because in~\cite{Ashesh2023-wtf}, the Lateral contextualization (LC) approach which incorporates context in a memory efficient way, worked well on Actin \vs Tubulin. 
Compared to \hdnusplit, our model is naturally at disadvantage because we have disabled the LC module but \hdnusplit uses it. 
We believe that increasing the patch size can help our \denoiSplit reduce the tiling artefacts. 

In general, we also find cases where \muSplit has retained some fine structures, albeit with noise, which the denoising based approaches have omitted from the prediction. We argue this to be a natural consequence of restricting the expressivity of latent spaces with KL divergence loss, which is pivotal for denoising. 

As joint denoising-Splitting is a new task, there is much that needs to be done. We humbly acknowledge the challenges mentioned above which we hope to tackle in our future works.  
\figrealNoise
\tabOtherDsets
\figFactinER
\figFactinMT
\figFactinCCPs

\figHDNERvsCCPs
\figHDNERvsMT
\figHDNCCPsvsMT
\figHDNFactinVsCCPs
\figHDNFactinVsMT
\figHDNFactinVsER
\tabMoreTasks
\figPaviaATN
\clearpage  

\end{document}